\documentclass[10pt, twocolumn, a4paper]{article}

\usepackage{amsmath,amssymb}
\usepackage[a4paper, top=2.0cm, bottom=1.5cm, left=1.5cm, right=1.5cm]{geometry}
\usepackage{helvet}
\usepackage{graphicx}
\usepackage{color}
\usepackage[colorlinks=true,bookmarks=false,citecolor=blue,urlcolor=blue]{hyperref}
\usepackage{physics}
\newcommand{\eps}{\varepsilon}      %Greek epsilon
\newcommand{\om}{\omega}      %Greek omega
\newcommand{\cra}[1]{\hat{a}^{\dag}_{#1}}  %creation operator a
\newcommand{\ana}[1]{\hat{a}_{#1}^{\vphantom{\dag}}}         %annihilation operator a

\title{Topological edge states of interacting photon pairs \\ emulated in a topolectrical circuit}

\author{Nikita A. Olekhno$^{1}$, Egor I. Kretov$^{1}$, Andrei A. Stepanenko$^{1}$, Polina A. Ivanova$^{1}$,\\
Vitaly V. Yaroshenko$^{1}$, Ekaterina M. Puhtina$^{1}$, Dmitry S. Filonov$^{2}$, Barbara Cappello$^{3}$,\\
Ladislau Matekovits$^{3}$, and Maxim A. Gorlach$^{1}$}

\begin{document}

\small

\twocolumn[
\begin{@twocolumnfalse}
\maketitle

{\it \small{
$^1$~Physics and Engineering Department, ITMO University, Saint Petersburg 197101, Russia\\
$^2$~Center for Photonics and 2D Materials, Moscow Institute of Physics and Technology, Dolgoprudny 141700, Russia
\\
$^3$~Department of Electronics and Telecommunications, Politecnico di Torino, I-10129 Torino, Italy
}}

\ \

\textbf{Topological physics opens up a plethora of exciting phenomena allowing to engineer disorder-robust unidirectional flows of light. Recent advances in topological protection of electromagnetic waves suggest that even richer functionalities can be achieved by realizing topological states of quantum light. This area, however, remains largely uncharted due to the number of experimental challenges. Here, we take an alternative route and design a classical structure based on topolectrical circuits which serves as a simulator of a quantum-optical one-dimensional system featuring the topological state of two photons induced by the effective photon-photon interaction. Employing the correspondence between the eigenstates of the original problem and circuit modes, we use the designed simulator to extract the frequencies of bulk and edge two-photon bound states and evaluate the topological invariant directly from the measurements. Furthermore, we perform a reconstruction of the two-photon probability distribution for the topological state associated with one of the circuit eigenmodes.}
%
%\textbf{Topolectrical circuitry has recently emerged as a simple yet very functional way of tabletop realization of topological physics. Employing this platform, we capture an intriguing behavior of two-photon topological states in quantum regime going beyond well-celebrated topological states of classical light. Besides their quantum nature, investigated two-photon states also exhibit an interaction-induced origin, elucidating the interplay between quantum entanglement and interactions. In our experiments, we extract the frequencies of bulk and edge two-photon bound states and evaluate the associated topological invariant directly from the measurements. To further enhance our findings, we perform a reconstruction of the two-photon probability distribution for the topological state associated with one of the circuit eigenmodes. Our results shed light onto the topological protection in interacting quantum-optical systems opening further perspectives for applications in quantum communication and quantum computations.}~\\~\\~\\
%\end{abstract}

\  \

\end{@twocolumnfalse}]

%_________________________________________Introduction___________________________________________
\section*{\small Introduction}

Recent years have brought a fast-paced development of topological physics in various systems ranging from traditional electronic setups~\cite{Kane-Mele,BHZ-Science,Hasan-Kane} or cold atom ensembles~\cite{Atala,Aidelsburger,Cooper-Spielman} to mechanical~\cite{Kane2014,Huber}, acoustic~\cite{Yang-Zhang} and electromagnetic~\cite{Wang-Soljacic,Lu2014, Lu2016, Khanikaev17, Ozawa_RMP} structures governed by classical wave equations. A remarkable feature of such systems is the presence of unidirectional topological states robust against disorder and exhibiting zero reflection at sharp bends. In this context, photonic topological states appear to be especially attractive offering an energy-efficient alternative to their electronic counterparts, allowing for easier scaling, control and manipulation and, ultimately, featuring a potential for on-chip integration~\cite{Hafezi:2013NatPhot,Mittal-2018,Blanco-Science}.

While topological states of classical light are relatively well-studied, topological states of quantum light are much less explored with only few first studies currently available~\cite{Mittal-2018,Blanco-Science,Barik, Tambasco, Wang-Pang}. At the same time,  topological states of quantum light can feature new exciting aspects of topological physics as, for instance, topological protection of biphoton correlations~\cite{Blanco-Science,Wang-Pang}. Further investigation of quantum-optical topological states may bring unexpected discoveries paving a way towards topologically protected quantum logic operations and quantum computations.

Nevertheless, the impact of interactions on quantum topological states remains largely uncharted, and first theoretical~\cite{Lee1,Gorlach-2017, DiLiberto,YKe,Valiente-19} and experimental~\cite{Roushan-Martinis,Roushan:2016NatPhys,Clark-Simon} studies demonstrate rich manifestations of topology in interacting systems. The major challenge here is the difficulty in implementation of sizeable nonlinearities manifested already at the two-photon level~\cite{Carusotto}.

\begin{figure*}[t]
    \centerline{\includegraphics[width=18cm]{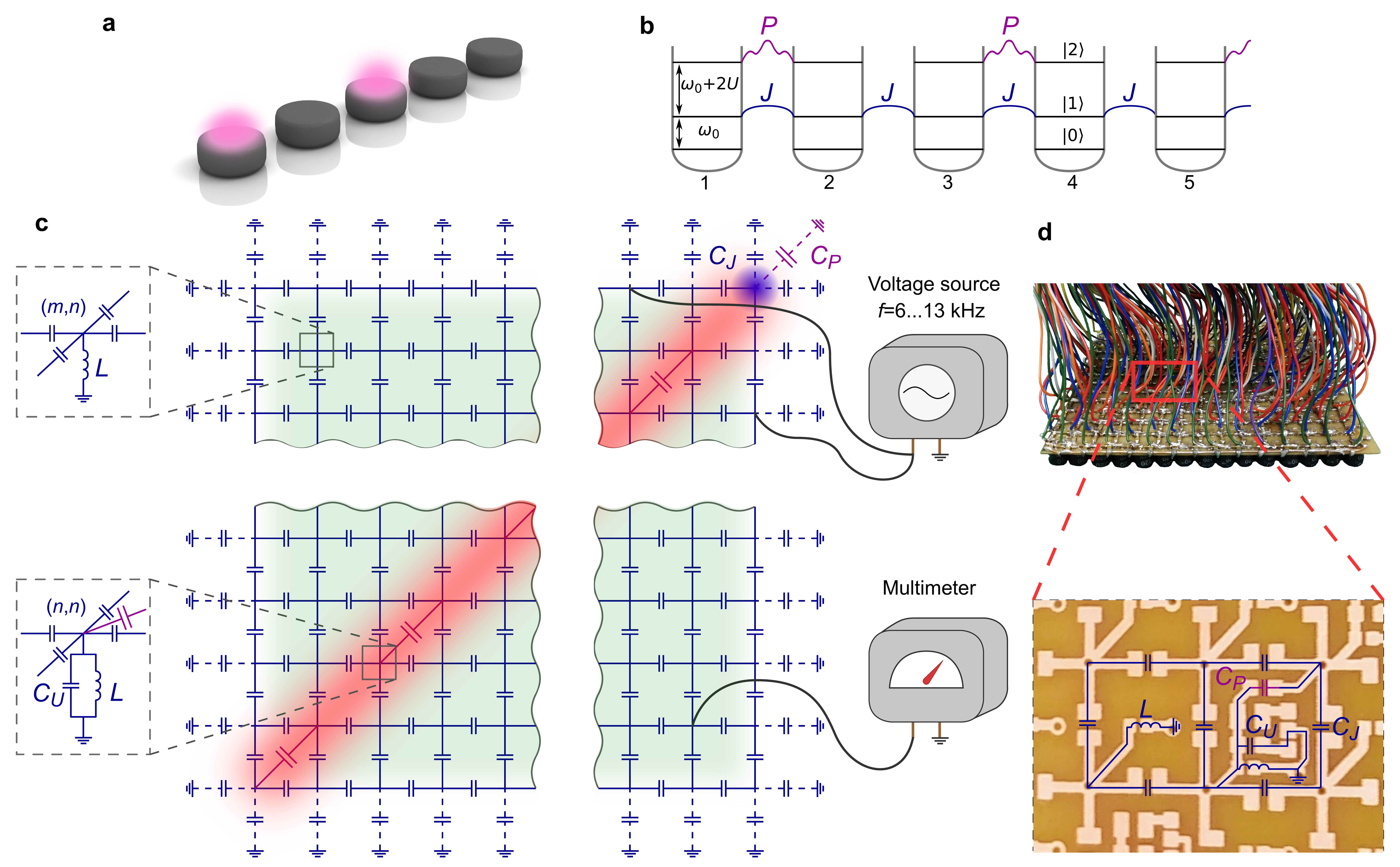}}
    \caption{\footnotesize {\bf Two-photon one-dimensional quantum problem and its topolectrical circuit simulator.} {\bf a}, Artistic view of two-photon excitations in the array of microresonators with tunneling couplings. The depicted state is $\cra{1}\,\cra{3}\,\ket{0}$. {\bf b}, Extended version of Bose-Hubbard model considered in the present article. Single-photon tunnelings $J$ are shown by blue solid lines, direct two-photon tunnelings $P$ are indicated by purple wavy lines. {\bf c}, Top view of the equivalent two-dimensional topolectrical circuit with a voltage at the site $(m,n)$ corresponding to probability amplitude $\beta_{mn}$ for one photon to be located at the $m^{\rm{th}}$ resonator of the array with another one located at the $n^{\rm{th}}$ resonator [cf. Eq.~\eqref{eq:psi}]. Colored regions show characteristic voltage patterns for two-photon scattering states (green), doublons (red), and doublon edge state (blue). External voltage source applied for the system excitation and voltmeter are shown to the right. Side view of the diagonal (lower inset) and off-diagonal (upper inset) sites of the topolectrical circuit, where   grounding elements are shown. {\bf d}, The photograph of experimental setup having the size of $15\times 15$ nodes. Inset shows the enlarged fragment of the circuit which includes two unit cells.}
    \label{fig:Network}
\end{figure*}

As a particular example of few-body interacting system, we consider a one-dimensional array of coupled nonlinear cavities described by the extended Bose-Hubbard model~\cite{Dutta} illustrated in Fig.~\ref{fig:Network}a. In this model, local photon-photon interactions give rise to exotic bound states of photon pairs persisting even in the case of repulsive nonlinearity~\cite{Mattis1986,Winkler} on which we focus from now on. Such repulsively bound pairs (doublons) were the subject of a series of recent theoretical~\cite{Valiente,Bello-2017, Salerno, Flach, Longhi, Marques2018} and experimental~\cite{Strohmaier2010,Preiss2015,Mukherjee,Greiner2017,Ye-Pan} studies. However, the observation of topological doublon edge states has remained elusive so far due to their absence in the standard Bose-Hubbard model with $P=0$~\cite{Flach}.  Very recently, several realizations of doublon edge states in various modified models have been suggested~\cite{Gorlach-2017, DiLiberto, Longhi, Gorlach-H-2017, DiLiberto-EPJ}, but their practical implementation still remained questionable.

Our present proposal relies on the extended version of Bose-Hubbard Hamiltonian
\begin{gather}
\hat{H} = \omega_0\sum_{m} \hat{n}_{m} -J \sum_{m}(\cra{m} \ana{m+1}+\cra{m+1} \ana{m})\label{eq:Hamiltonian}\\
+U \sum_{m} \hat{n}_{m}(\hat{n}_{m}-1)+\frac{P}{2}\sum_{m} (\cra{2m-1}\cra{2m-1}\ana{2m}\ana{2m} +\text{H.c.})\:,\notag
\end{gather}
illustrated schematically in Fig.~\ref{fig:Network}b. Here, the summation is performed over all resonators in the array enumerated with the index $m$, H.c. denotes the Hermitian conjugate of the term to the left, $\cra{m}$ and $\ana{m}$ are the creation and annihilation operators for the photon at the $m^{\rm{th}}$ resonator, and $\hat{n}_{m} = \cra{m}\ana{m}$ is the photon number operator. The first term in the above Hamiltonian defines the energy of non-interacting photons in the resonator and has a trivial effect on the spectrum of photon pairs shifting it by $2\,\om_0$. Therefore, for simplicity we use $2\,\omega_0$ as an energy reference for the two-photon excitations and omit the corresponding term. The second term of the Hamiltonian describes single-photon tunneling  between $m^{\rm{th}}$ and $(m+1)^{\rm{st}}$ resonators in the array. Taken together, these two terms describe the linear array of identical cavities with the lowest eigenfrequency $\omega_0$ and the tunneling coupling $J$ between the nearest neighbors. The latter two terms of the Hamiltonian Eq.~\eqref{eq:Hamiltonian} describe effective photon-photon interactions mediated by the nonlinearity of the medium including an on-site photon-photon interaction $\propto U$ and a direct two-photon hopping $\propto P$, respectively. 

A key property of this system is the emergence of bound two-photon topological  states along with the trivial single-photon excitations. Hence, the topological order in this system is facilitated by the effective photon-photon interaction which provides the simplest example of interaction-induced topological states of quantum light. Furthermore, in the strong interaction limit $U\gg J$ doublon excitations are described by the effective Su-Schrieffer-Heeger Hamiltonian~\cite{Su} which is a paradigmatic one-dimensional topological model (see the analysis below and the Supplementary Note~1).

To overcome the difficulties with the direct experimental implementation of our model Eq.~\eqref{eq:Hamiltonian} and to bridge the gap between quantum-optical topological states and physics of interacting systems, we adopt the concept of topolectrical circuits~\cite{Ningyuan,Albert,Imhof, Topolectrical, Hadad-Nature,Chong-NC, Hu-NComm, Serra-Garcia-Huber,Helbig,Hofmann} applying them to emulate an interacting two-body problem in one dimension. As detailed below, this  approach is based on mathematically rigorous mapping of quantum two-body problem onto the classical setup of higher dimensionality, and this correspondence renders the topolectrical platform a powerful tool to study topological states of interacting photons.

%_________________________________________Results_______________________________________________
\section*{\small Results}

%____________________________Circuit_____________________________
{\bf Topolectrical circuit realization and types of two-photon excitations}. Two-photon solutions to the stationary Schr{\"o}dinger equation $\hat{H}\,\ket{\psi} = \eps\,\ket{\psi}$ with the Hamiltonian Eq.~\eqref{eq:Hamiltonian} can be searched in the form
\begin{equation}\label{eq:psi}
    \ket{\psi} = \frac{1}{\sqrt{2}}\,\sum_{m,n=1}^{N}\beta_{mn}\,\hat{a}^{\dag}_{m}\hat{a}^{\dag}_{n}\ket{0}\:,
\end{equation}
where $\ket{0}$ is the vacuum state and $N$ is the total number of resonators in the array. Superposition coefficients $\beta_{mn}$ characterize the probability amplitude for one photon to be present at site $m$ with the other one located at site $n$. Due to bosonic nature of the problem, superposition coefficients are symmetric: $\beta_{mn}=\beta_{nm}$.

The eigenvalue problem with the wave function Eq.~\eqref{eq:psi} and the Hamiltonian Eq.~\eqref{eq:Hamiltonian} yields a linear system of equations with respect to the unknown coefficients $\beta_{mn}$ which can be written as
\begin{equation}\label{eq:eig-equation}
\sum_{m',n'}\,\left[H_{mn,m'n'}-\eps\,\delta_{mn,m'n'}\right]\,\beta_{m'n'}=0\:.
\end{equation}
Off-diagonal elements of the matrix $H_{mn,m'n'}$ describe either single-photon or two-photon tunneling coupling between the sites $(m,n)$ and $(m',n')$, whereas the diagonal entries $H_{mn,mn}-\eps$ include resonator frequency  detuning $U$ and energy variable $\eps$. 

To explore the physics of repulsively bound photon pairs and their edge states, we notice that the original one-dimensional two-particle problem is described by the same discrete wave equation as a two-dimensional tight-binding system. In such setup, vertical and horizontal links represent the hopping amplitudes for the first and second photons, additional links coupling the sites $(2n-1,2n-1)$ and $(2n,2n)$ emulate direct two-photon tunneling process, whereas eigenfrequency shift for the diagonal sites $(n,n)$ represents on-site interactions $U$ in the original quantum-optical problem [Fig.~\ref{fig:Network}a].

The described tight-binding system can be readily implemented experimentally using arrays of coupled waveguides~\cite{Rechtsman}, coupled ring resonators~\cite{Hafezi-2011,Hafezi:2013NatPhot} or even LC circuits~\cite{Imhof}. Still, any such classical two-dimensional realization emulates a pair of distinguishable particles. In the latter case, the dynamics is governed by the first-quantized Hamiltonian
\begin{equation}\label{eq:HamiltonianFirst}
\begin{split}
\hat{H}=-J\,\sum\limits_{n}\,\left(\ket{x_{n}}\bra{x_{n+1}}+\ket{y_{n}}\bra{y_{n+1}}+\text{H.c.}\right)\\
+2U\,\sum\limits_{n}\,\ket{x_{n},y_{n}}\bra{x_{n},y_{n}}\\
+P\,\sum\limits_{n}\,\left(\ket{x_{n},y_{n}}\bra{x_{n+1},y_{n+1}}+\text{H.c.}\right)\:,
\end{split}
\end{equation}
while the wave function in the first-quantized form reads:
\begin{equation}
\ket{\psi}=\sum\limits_{m,n}\,\beta_{mn}\,\ket{x_{m},y_{n}}\:,
\end{equation}
where $1/\sqrt{2}\,\left[\hat{a}^{\dag}_{n}\right]^2\,\ket{0}$ corresponds to $\ket{x_{n},y_{n}}$, and $\cra{m}\,\cra{n}\,\ket{0}$ corresponds to the combination $1/\sqrt{2}\,\left(\ket{x_{m},y_{n}}+\ket{x_{n},y_{m}}\right)$ ($m\not=n$). Note that the states with bosonic symmetry can be emulated exciting this setup symmetrically with respect to the diagonal.

Choosing an appropriate platform, we aim not only to observe the excitation of doublon edge state, but also to reconstruct the associated probability distributions for the eigenmodes and to extract the topological invariant for bulk doublon bands directly from the experiment. Reaching this goal implies extensive measurements of field amplitudes at all relevant sites of the system for different excitation scenarios. Based on this, we choose the most accessible platform of topolectrical circuits, for which the node potentials $\varphi_{mn}$ correspond to the $\beta_{mn}$ coefficients in tight-binding equations.

To design the desired two-dimensional topolectrical circuit, we combine the first and the second Kirchoff's rules into the matrix equation
\begin{equation}\label{eq:Kirchhoff}
    \sum_{m',n'}\, Y_{mn,m'n'}\,\varphi_{m'n'}=0\:.
\end{equation}
Here, every composite index $mn$ labels one site of the two-dimensional lattice having the coordinates $(m, n)$. Off-diagonal entries of the matrix $Y_{mn,m'n'}$ are equal to the admittances of the elements directly connecting the sites $(m,n)$ and $(m',n')$ in the circuit, while the diagonal elements are defined as 
\begin{equation}\label{eq:DiagSigma}
Y_{mn,mn}=-Y_{mn}^{(\rm{g})}-\sum_{(m',n') \neq (m,n)} Y_{mn,m'n'}\:,
\end{equation}
where $Y_{mn}^{(\rm{g})}$ is the admittance of the element connecting site $(m,n)$ to the ground. Comparing Eqs.~\eqref{eq:eig-equation} and \eqref{eq:Kirchhoff}, we immediately recover that off-diagonal entries of the admittance matrix correspond to tunneling couplings $J$ or $P$ in the initial tight-binding model, while the diagonal elements are associated with the resonator detuning $U$. Additionally, to realize the desired tight-binding model Eq.~\eqref{eq:eig-equation}, the lack of neighbors for the edge or corner sites of a topolectrical circuit evident from Eq.~\eqref{eq:DiagSigma} should be compensated by the proper adjustment of admittance $Y_{mn}^{(\rm{g})}$. Further analysis carried on in Supplementary Note~3 provides the following identification of tight-binding parameters in terms of circuit elements:
\begin{equation}\label{eq:ParamId}
    J = 1,\mspace{8mu} U = \frac{C_P+C_U}{2\,C_J}, \mspace{8mu} P = -\frac{C_P}{C_J},
\end{equation}
whereas the ``energy'' eigenvalue is inversely proportional to the frequency $f$
\begin{equation}
    \varepsilon = \frac{f_0^2}{f^2} - 4, \quad f_0^2 = \frac{1}{4\,\pi^2\,LC_J}.
\end{equation}
The scheme of the designed circuit is shown in Fig.~\ref{fig:Network}c, while the photograph of the experimental sample is provided in Fig.~\ref{fig:Network}d. For the experimental setup, the chosen element values result in $U=7.09$, $P=-4.18$ which ensure that doublon bands are well-separated from the continuum of scattering states and hence can be reliably detected.

\begin{figure*}[t]
    \centerline{\includegraphics[height=14cm]{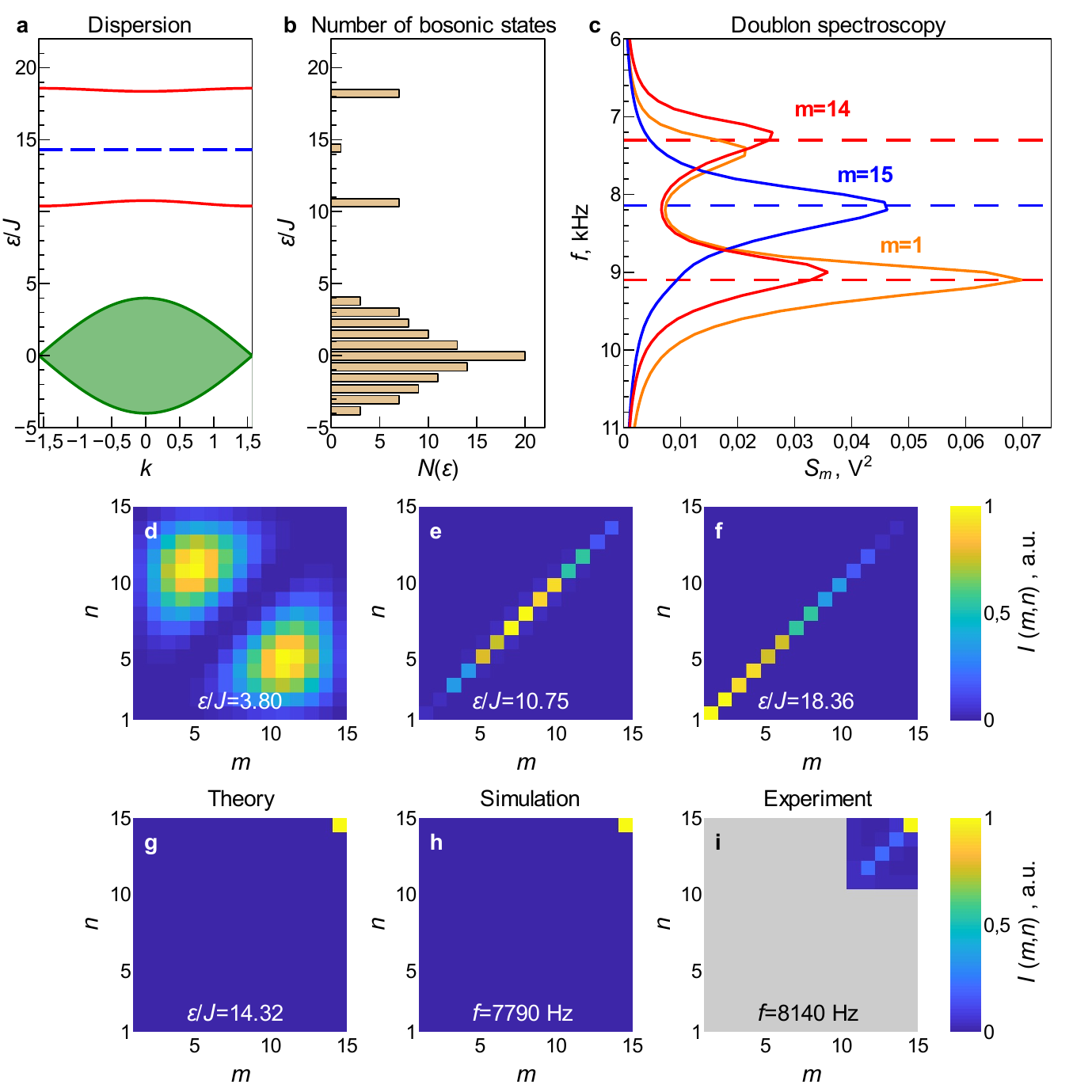}}
    \caption{\footnotesize {\bf Theoretical studies and experimental emulation of two-photon excitations.} {\bf a}, Dispersion of two-photon eigenmodes calculated from the tight-binding model. Two red solid curves correspond to doublons, horizontal dashed line between them indicates the energy of the doublon edge state, and the shaded area at the bottom shows the continuum of two-photon scattering states. {\bf b}, Number of bosonic states in the tight-binding system of size $15 \times 15$ sites. {\bf c}, Doublon spectroscopy with quantity $S_{m}(f)$ [cf. Eq.~\eqref{eq:S}] determined from experimental data and plotted as a function of driving frequency. The feeding points $(1,1)$, $(14,14)$ and $(15,15)$ are labelled with corresponding values of $m=1,14,15$ and are shown by orange, red and blue solid lines, respectively. Characteristic peaks in the spectrum correspond to doublon modes. {\bf d-g}, Two-photon probability distributions $|\beta_{mn}|^2$ for the eigenmodes of $15 \times 15$ lattice described by the Hamiltonian Eq.~\eqref{eq:Hamiltonian} in the absence of disorder. {\bf h}, Eigenmode tomography for a circuit of $15 \times 15$ nodes simulated taking into account the disorder in the element values as well as Ohmic losses (see Methods section for details). {\bf i}, Experimental implementation of the eigenmode reconstruction for the doublon edge state.}
    \label{fig:Modes}
\end{figure*}

Tight-binding calculations suggest that the eigenmodes supported by the designed structure can be classified into three types with the spectrum shown in Fig.~\ref{fig:Modes}a. The lower series of bands shown by green is symmetric with respect to zero energy and corresponds to the two-photon scattering states. In the infinite-size limit, the photons in such states have energies in the continuous spectrum given by the sum of single-photon continuum energies. The typical probability distribution for the scattering state is depicted in Fig.~\ref{fig:Modes}d.

Two bands present at higher energies correspond to doublons. Probability distributions depicted in Fig.\ref{fig:Modes}e,f suggest that the two photons most likely share the same resonator being free to move along the entire array. As a consequence, $\beta_{nn}$ coefficients are the dominant ones in the expansion of the two-photon wave function.

Finally, the gap between two bulk doublon modes is occupied by the doublon edge state with localization illustrated in  Fig.~\ref{fig:Modes}g. In the limit $U\gg J$ and $|P|\gg J$ the energy of doublon edge state scales as $2U$, whereas the splitting between the two bands is approximately $2|P|$ (see Supplementary Note~1 for details). Thus, one can broadly tune the system behavior by varying the parameters $U$ and $P$ defined via the capacitances of circuit elements, Eq.~\eqref{eq:ParamId}.

In what follows, we focus on doublon states with a special emphasis on the doublon edge state which we prove to be topological. At first glance, such zero-dimensional localized state in a two-dimensional circuit may seem similar to higher-order topological states which have recently been predicted and observed in various systems~\cite{Benalcazar2017, Benalcazar2017b, Schindler2018,Nori} ranging from solid state~\cite{Schindler2018b} to photonics~\cite{Mittal-NP}. However, despite the seeming similarity, our proposal accesses completely different physics associated with quantum-optical topological states in interacting two-particle models emulated with the help of classical system of higher dimensionality. Quite remarkably, even in such situation topolectrical circuits provide a possibility to probe frequencies and probability distributions of doublon bulk and edge states, giving valuable information on the original 1D quantum problem.

%______________________Measurements_______________________
{\bf Experimental results}. To determine the spectral positions of doublon modes in the experiment, we apply voltage to one of the diagonal sites of the circuit, $(m,m)$, keeping the track of potentials $\varphi_{nn}^{mm}$ at all diagonal sites $(n, n)$. Next we construct the quantity~\cite{Gorlach-2018}
\begin{equation}
    S_{m}(f)=\sum_{n}|\varphi_{nn}^{mm}|^2,
    \label{eq:S}
\end{equation}
which is evaluated as a function of driving frequency $f$. While scattering states feature zero overlap with the diagonal sites, doublon modes are characterized by the voltage maxima at the diagonal. Therefore, chosen excitation scenario selects just doublon modes, and their frequencies can be immediately associated with the characteristic peaks in the spectrum of $S_m$ clearly seen in Fig.~\ref{fig:Modes}c. Thus, $S_m(f)$ appears to be more convenient quantity compared to circuit impedance (Supplementary Note~7), since it filters out the contribution of the two-photon scattering states.

Experimental spectrum of $S_m$ plotted for $m=1,14$ and $15$ in Fig.~\ref{fig:Modes}c features two characteristic peaks once the voltage is applied to the sites $(1,1)$ or $(14,14)$, whereas feeding of the site $(15,15)$ results in a single peak. Based on our analytical model (see Methods section), we associate the former two peaks with bulk doublon modes, whereas the latter peak provides a clear signature of the doublon edge state. The respective eigenfrequencies are approximately $7.28$~kHz and $9.08$~kHz for bulk doublon modes and $8.14$~kHz for the doublon edge state. A significant broadening of the peaks observed in Fig.~\ref{fig:Modes}c is caused by the combination of such factors as Ohmic losses inevitable in a real  system and the dispersion of actual values of the circuit elements (see Methods for details).

The observed characteristic peaks in $S_m(f)$ provide an indirect evidence of bulk and edge doublon states. More information can be retrieved by measuring the voltage distribution at the nodes of the system at a  given frequency. This distribution, however, depends strongly on the choice of the node $(m,m)$ at which the voltage is applied. In particular, exciting the system at $(15,15)$ node, we recover the voltage pattern with a maximum at the point of excitation which can be explained either as doublon edge state or just as a trivial defect at the corner of the system.

To provide a clear evidence of doublon states independent of the choice of the feeding point, we have performed the reconstruction of voltage distribution for the doublon edge state eigenmode at the characteristic frequency of the corresponding peak. Such eigenmode tomography described in detail in Supplementary Note~2 employs the following steps. Throughout the whole procedure the frequency of excitation, $f$, and the external voltage are fixed. For some symmetric choice of the feeding points, e.g. $(m,n)$ and $(n,m)$, we measure full voltage distribution $\varphi_{m'n'}^{mn}$ at all nodes $(m',n')$ of the circuit and then evaluate the quantity
\begin{equation}\label{eq:Tomography1}
\mathfrak{I}(m,n)=\sum\limits_{m',n'}\,|\varphi_{m'n'}^{mn}|^2\:.
\end{equation}
Performing this step for various symmetric choices of the feeding points, we get an entire array of values for $\mathfrak{I}$ which is now considered as a discrete function of $m$ and $n$ coordinates. Finally, we depict this function on a colorplot  Fig.~\ref{fig:Modes}i and observe a good agreement with the eigenmode distribution evaluated from the tight-binding model, Fig.~\ref{fig:Modes}g, as well as with the numerical solution of Kirchhoff's equations, Fig.~\ref{fig:Modes}h.

As shown in Supplementary Notes~2 and 4, the procedure outlined above reproduces the quantity
\begin{equation}\label{eq:Tomography2}
    \mathfrak{I}(m,n) \propto \sum_k \frac{|\varphi_{mn}^{(k)}|^2}{(f - f_k)^2 + \gamma^2}\:,
\end{equation}
where $\varphi_{mn}^{(k)}$ is the potential at site $(m,n)$ for the eigenmode with frequency $f_k$, and $\gamma$ is the effective dissipation rate related to Ohmic resistance of the circuit elements. Hence, $\mathfrak{I}(m,n)$ performs a sum over all eigenmodes with a Lorentzian-type weighting factor having a sharp maximum for the eigenmode with frequency $f_k$ matching the excitation frequency $f$.

Furthermore, Eq.~\eqref{eq:Tomography2} reveals that eigenmode tomography works exceptionally well when the spectral separation of the eigenmode (or a band of eigenmodes with a similar intensity pattern) from the rest of the modes exceeds the effective dissipation rate which is true for the doublon edge state, Fig.~\ref{fig:Modes}i. As a consequence, all key features of the corresponding state are well-reproduced with only slight distortions present. On the other hand, all modes of the scattering continuum feature much worse results of eigenmode tomography due to relatively large density of  two-photon scattering states [Fig.~\ref{fig:Modes}b] and quite different field distributions for the different modes. At any case, such scattering states are not especially interesting since they feature neither topological protection, nor pronounced effects of interaction.

\begin{figure}[t]
    \centerline{\includegraphics[width=8cm]{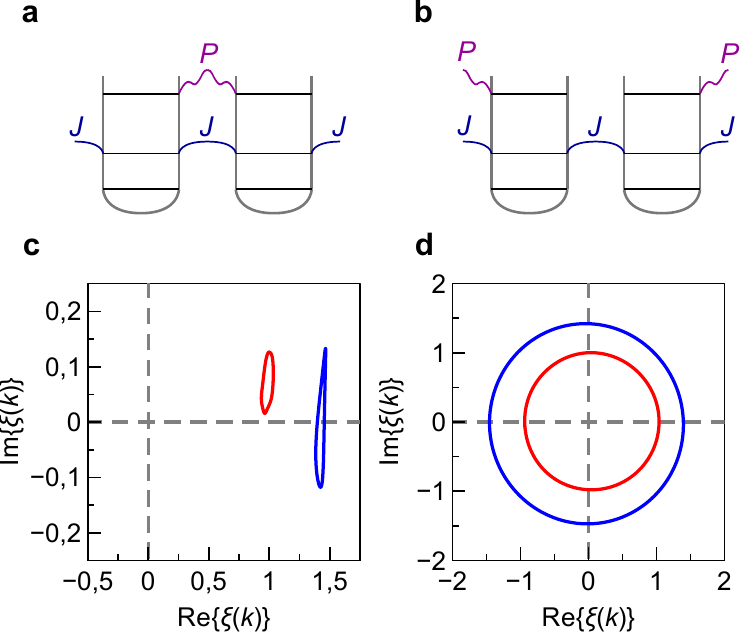}}
    \caption{\footnotesize {\bf Topological characterization of doublon modes.} {\bf a-b}, Two possible choices of the unit cell. {\bf c-d}, The ratio of sublattice voltages $U_{\rm A}(k)$ and $U_{\rm B}(k)$, $\xi(k)=U_{\rm A}(k)/U_{\rm B}(k)$ plotted on a complex plane for the wave number $k$ varied over the entire Brillouin zone. Blue and red lines correspond to numerical simulation and experiment, respectively. The winding numbers $W=0$ (panel {\bf c}) and $W=1$ (panel {\bf d}) correspond to the unit cell choiches shown in panels {\bf a} and {\bf b}. The voltage distribution is measured for the system excited at the node $(7,7)$.}
    \label{fig:Topology}
\end{figure}

Experimental results of eigenmode tomography presented in Fig.~\ref{fig:Modes}i demonstrate good agreement with our theoretical expectations. Note also that while doublon edge state exhibits some hybridization with bulk doublons [Fig.~\ref{fig:Modes}e,f], it does not mix with the scattering states, which is a consequence of non-overlapping spatial distributions of these modes and considerable spectral separation between them.

Having reconstructed the profile of the doublon edge eigenmode directly from the experimental data [Fig.~\ref{fig:Modes}i], we now turn to the discussion of its topological origin. Since the effective interaction strength $U$ is considerably larger than $J$, doublons are mostly localized at the diagonal of the 2D structure being effectively one-dimensional. Hence the standard technique based on winding number calculation~\cite{Ryu} can be applied.

Winding number in Su-Schrieffer-Heeger-type models is determined by plotting the ratio of voltages at the two sublattices, $U_{\rm A}(k)$ and $U_{\rm B}(k)$, on the complex plane for a particular Bloch eigenmode, the topological scenario being characterized by the curve enclosing the coordinate origin. The winding number $W$ is defined as the number of revolutions of this curve around the coordinate origin. Note also that the winding number depends on the unit cell choice and therefore to reveal the topological edge state, the choice of the unit cell should be consistent with the array termination.

In the experiment, we set the excitation frequency to that of the upper bulk doublon band, $f=9550$~Hz, and measure the distribution of voltages at the diagonal keeping the track of their phase. Recovering Bloch eigenmodes via Fourier transform and extracting sublattice voltages as outlined in Methods section, we finally get winding number graphs depicted in Fig.~\ref{fig:Topology}. Again, in a close agreement with our theoretical predictions, we recover that the unit cell choice without $P$ link inside gives rise to the nontrivial winding thus proving the topological origin of the observed doublon edge state. Note also that another choice of the unit cell with $P$ link inside yields zero winding number proving the absence of topological doublon edge state at the opposite $(1,1)$ corner of the array.

%_________________________________________Discussion_____________________________________________
\section*{\small Discussion}

To summarize, we have implemented a two-dimensional topolectrical circuit serving as an analogue simulator for quantum  one-dimensional two-particle interacting problem. Using the exact mapping of the initial quantum-optical system of two entangled interacting photons onto the classical setup, we have been able to simulate bulk and edge states of repulsively bound photon pairs.

Examining various excitation scenarios of the designed 2D circuit, we have not only reconstructed the doublon edge mode directly from experimental data, but also extracted the associated winding number from the measurements thus providing a rigorous proof of topological doublon edge state existence. Note that the possibility to conduct such extensive measurements is a distinctive feature of topolectrical platform, whereas the implementation of the  same  protocols for optical systems remains highly challenging.

The same approach can be used to emulate the dynamics of $N$ interacting particles in one spatial dimension. However, the dimensionality of the setup needed for that purpose should be equal to $N$ and therefore this emulation approach is meaningful only for reasonably low $N$ and only for the Hamiltonians conserving the number of particles.

Thus, we believe that our emulation of two-photon topological states induced by interactions uncovers new intriguing aspects of topological physics in interacting systems providing further insights into topological protection of quantum light.

%_________________________________________Methods______________________________________________
\section*{\small Methods}

{\scriptsize

\textbf{Tight binding equations}

The eigenmodes of quantum-optical problem under study correspond to the eigenvectors of the Hamiltonian Eq.~\eqref{eq:Hamiltonian}. Since the Hamiltonian commutes with the operator $\sum_m\,\hat{n}_m$, the total number of photons is conserved. Hence, the wave function of arbitrary two-photon excitation can be represented in the form Eq.~\eqref{eq:psi}. Combining Eqs.~\eqref{eq:Hamiltonian} and \eqref{eq:psi} into an eigenvalue equation, we obtain the linear system of equations with respect to the unknown coefficients $\beta_{mn}$ which provide a probability amplitude for two photons to be present in $m^{\rm{th}}$ and $n^{\rm{th}}$ cavities: 
\begin{gather}
  (\eps-2U)\beta_{2m,2m}=-2J\,\left[\beta_{2m+1,2m}+\beta_{2m,2m-1}\right] + P \beta_{2m-1,2m-1}\:,\label{BetaEqs1}\\
  (\eps-2U)\beta_{2m-1,2m-1} =-2J\,\left[\beta_{2m,2m-1}+\beta_{2m-1,2m-2}\right]+ P \beta_{2m,2m}\:,\label{BetaEqs2}\\
  \eps\beta_{m,n} = - J\,\left[\beta_{m+1,n} +\beta_{m-1,n}+\beta_{m,n+1}+\beta_{m,n-1}\right]\:,\mspace{8mu} (m\not=n)\label{BetaEqs3}
\end{gather}
where $\eps$ is the eigenmode energy, and $\beta_{mn}=\beta_{nm}$ due to the bosonic nature of the problem. The dispersion of two-photon modes and associated probability distributions plotted in the main text are calculated by solving the eigenvalue problem Eqs.~\eqref{BetaEqs1}-\eqref{BetaEqs3}.

\ \

\textbf{Excitation of the modes in the topolectrical circuit}

The eigenmodes of the circuit are found as the solutions to the eigenvalue problem Eq.~\eqref{eq:Kirchhoff}. As further discussed in Supplementary Note~3, this eigenvalue problem has a one-to-one correspondence with the initial tight-binding system.

To describe the response of the circuit to the external excitation, we combine the first and the second Kirchhoff's rules to derive the equation with respect to the unknown potentials at the sites of the system. Excitation of the system is described as an external current injected symmetrically into $(p,q)$ and $(q,p)$ nodes of the circuit:
\begin{equation}\label{eq:CircuitExcitation}
    \sum_{m',n'}\, Y_{mn,m'n'}\,\varphi_{m'n'}^{pq}=\frac{I_0}{2}\,\left[\delta_{mp}\delta_{nq}+\delta_{mq}\delta_{np}\right]\:.
\end{equation}
Inverting the matrix of admittances which enters Eq.~\eqref{eq:CircuitExcitation}, we immediately obtain the distribution of voltages at the nodes. To ensure reasonable results at resonances of the circuit, Ohmic losses should be properly taken into account. As further discussed in Supplementary Note~4, a topolectrical circuit with weak losses can be mapped onto the general driven-dissipative model studied in Ref.~\cite{Gorlach-2018}.

Other important features of our topolectrical implementation include the negative sign of $P$ as guaranteed by Eq.~\eqref{eq:ParamId}. The designed system is mounted on RF51 substrate and includes the following elements illustrated in Fig.~\ref{fig:Network}c,d:  $L = 22.77\,{\rm \mu H}$, $C_J = 1\,{\rm \mu F}$, $C_U = 10\,{\rm \mu F}$ and $C_P = 4.2\,{\rm \mu F}$. Ohmic losses are introduced as a resistance attached in series to the inductance or capacitance. At frequencies of interest ($6.0<f<13.0$~kHz) such parasitic resistances are estimated as $R^{L}=0.026$~Ohm, $R^{C}_{J}=0.2$~Ohm, $R^{C}_{U}=0.1$~Ohm and $R^{C}_{P}=0.4$~Ohm based on specifications of the  elements. Additionally, we also incorporate the effects of disorder by reconstructing the entire map of the elements placed onto the fabricated circuit (Supplementary Note~5). The associated value distributions feature a broadening of up to 2\%. Quite importantly, at frequencies of interest the parasitic resistance $R^L$ plays the major role. For that reason, designing our sample, we have chosen high-quality inductance coils with a sufficiently low dispersion in the values of inductance. There are other sources of disorder in the fabricated system as well, including non-ideal contacts between soldered elements, parasitic inductances of tracks and capacitances created by the circuit board itself, contacts between the board and the external measurement devices.

\ \

\textbf{Winding number}

To prove the topological nature of the doublon edge state, we apply the standard technique based on winding number evaluation~\cite{Ryu}. To calculate the winding number for a chiral-symmetric system, one has to convert the Bloch Hamiltonian to the off-diagonal form
\begin{equation}\label{eq:ChiralHam}
\hat{H}(k)=\begin{pmatrix}
0 & \hat{Q}(k)\\
\hat{Q}^\dag(k) & 0
\end{pmatrix}
\end{equation}
and then plot the curve for $\det\hat{Q}(k)$ on the complex plane with Bloch wave number $k$ spanning the entire Brillouin zone. The winding number is determined as a number of revolutions of the curve around the coordinate origin.

In the case of the Su-Schrieffer-Heeger model, Bloch Hamiltonian has the dimensions $2\times 2$ taking the form Eq.~\eqref{eq:ChiralHam} in the basis constructed from the modes of isolated resonators. Therefore, the scalar function $Q(k)$ is proportional to the ratio of  voltages at the two sublattices (even sites and odd sites) which provides a relatively simple way to extract the winding number from experiment, as has been done e.g. in Ref.~\cite{Rosenthal}.

In our case, chiral symmetry of doublon bands is only an approximation which holds in the limit $U\gg J$ only (see Supplementary Note~1 for details). In such a situation, one can neglect the mixing between the scattering continuum and doublon bands using an effective $2\times 2$ doublon Bloch Hamiltonian. Furthermore, in such a case doublons are mostly localized at the diagonal, and therefore only the diagonal nodes of the circuit have to be examined.

Accordingly, we set the frequency of excitation to that of higher-frequency bulk doublon band, $f=9550$~Hz, and measure voltage distribution at all diagonal nodes with an oscilloscope with a driving voltage applied to $(7,7)$ node of the circuit. In the corresponding numerical simulation we set excitation frequency to $f=8730$~Hz to match the frequency of doublon peak as it appears in the simulations. The relative phase of voltages at the nodes is determined by measuring their time dependence and fitting the dependence by the sinusoidal function with the unknown phase. To get the results for a single Bloch mode, we perform a Fourier transform of the obtained voltages as follows:
\begin{gather}
U_{\rm A}(k)=\sum_n U_{\rm A}(n)\,e^{-ikn}\:,\mspace{8mu} U_{\rm B}(k)=\sum_n U_{\rm B}(n)\,e^{-ik(n-1)}\:,
\end{gather}
where the indices $A$ and $B$ refer to the two different sublattices (even and odd sites). Next, we plot the ratio $\xi(k)=U_{\rm A}(k)/U_{\rm B}(k)$ on the complex plane for different choices of the unit cell with Bloch wave number spanning the range $[-\pi,\pi]$.

}

%______________________________________Data_availability___________________________________________
\subsection*{\small Data availability}
The data that support the findings of this study are available from the corresponding author upon request.

\subsection*{\small Acknowledgments}
We acknowledge valuable discussions with Alexander Poddubny, Alexander Khanikaev, Alexey Slobozhanyuk and Evgenii Svechnikov. N.O. thanks Sergey Tarasenko for stimulating discussions on topological physics. Theoretical models were supported by the Russian Foundation for Basic Research (grant No.~18-29-20037), experimental studies were supported by the Russian Science Foundation (grant No.~16-19-10538). N.O. and M.G. acknowledge partial support by the Foundation for the Advancement of Theoretical Physics and Mathematics ``Basis''.

%______________________________________Author_contributions_______________________________________
\subsection*{\small Author contributions}
M.G. conceived the idea and supervised the project. A.S., N.O. and M.G. worked out the theoretical models. N.O. and B.C. performed numerical simulations. N.O., M.G. and D.F. developed the circuit model and designed the experiment. E.K., D.F., V.Y., P.I. and E.P. fabricated the experimental setup. E.K., V.Y, N.O., P.I. and E.P. conducted the experiments. N.O., E.K. and B.C. supervised by L.M. performed post processing of the experimental data. N.O. and M.G. prepared the manuscript. All authors contributed extensively to the discussion of the results.

\subsection*{\small Competing interests}
The authors declare that they have no competing interests.

\subsection*{\small Additional information}
Correspondence and requests for materials should be addressed to M.G. (email: m.gorlach@metalab.ifmo.ru).

%________________________________________Bibliography____________________________________________
{\scriptsize

%\bibliographystyle{naturemag}
%\bibliography{TopologicalLib}
}

\end{document}

\renewcommand\refname{Supplementary References}
\maketitle

\newpage

\small{
\tableofcontents
}

\newpage

%%%%%%%%%%%%%%%%%%%%%%%%%%%%%%%%%%%%%%%%%%%%%%%%%%%%%%%%%%%%%%%%%%%%%%%%%%
\section{Supplementary Note 1~-- Analysis of the strong interaction limit $U\gg J$: Su-Schrieffer-Heeger model for doublons}\label{sec:SSHmap}

To provide a simple proof of the topological origin of our model, we analyze the strong interaction limit, when $U\gg J$. In such a situation, doublon bands are well-separated from the continuum of two-photon scattering states so that the mixing between doublons and scattering states is negligible. As such, we introduce an effective doublon Hamiltonian which captures the dynamics of doublons excluding other redundant degrees of freedom.

To derive the effective doublon Hamiltonian, we start from the eigenvalue equations provided in the article main text (Methods section):
\begin{gather}
  (\eps-2U)\beta_{2m,2m}=-2J\,\left[\beta_{2m+1,2m}+\beta_{2m,2m-1}\right] + P \beta_{2m-1,2m-1}\:,\label{BetaEq1}\\
  (\eps-2U)\beta_{2m+1,2m+1} =-2J\,\left[\beta_{2m+2,2m+1}+\beta_{2m+1,2m}\right]+ P \beta_{2m+2,2m+2}\:,\label{BetaEq2}\\
  \eps\beta_{m,n} = - J\,\left[\beta_{m+1,n} +\beta_{m-1,n}+\beta_{m,n+1}+\beta_{m,n-1}\right]\:,\mspace{8mu} (m\not=n)\label{BetaEq3}
\end{gather}
In the limiting case $U\gg J$, $\beta_{mm}$ are the dominant coefficients of the doublon wave function with the rest of coefficients $\beta_{m+s,m}$ decaying with the index $s$. Therefore, we neglect all terms proportional to $\beta_{m+2,m}$, $\beta_{m+3,m}$, etc. in Supplementary Equations~\eqref{BetaEq1}-\eqref{BetaEq3}, treating terms proportional to $\beta_{m+1,m}$ as a perturbation.

Using the approximation $\beta_{m+1,m}\approx -\left(\beta_{mm}+\beta_{m+1,m+1}\right)/(2\,U)$, we derive the following approximate eigenvalue equation for doublon bands:
\begin{gather}
(\eps-2U-2j)\beta_{2m,2m}=(j+P) \beta_{2m-1,2m-1} + j\,\beta_{2m+1,2m+1}\:,\label{systeff1}\\
(\eps-2U-2j)\beta_{2m+1,2m+1} =j\,\beta_{2m,2m} + (j+P)\,\beta_{2m+2,2m+2}\:.\label{systeff2}
\end{gather}
The equations for $\beta_{11}$ and $\beta_{NN}$ corresponding to the physical edges of the array are modified if compared to the bulk sites:
\begin{gather}
(\eps-2U-j)\beta_{11} =(j+P)\,\beta_{22}\:,\label{boundeff1}\\
(\eps-2U-j)\beta_{NN} = j\,\beta_{N-1,N-1}\:,\label{boundeff2}
\end{gather}
where $j = J^2/U$ is the effective doublon hopping rate associated with two consecutive single-particle tunnelings to the neighboring cavity. Here, the array length $N$ is assumed to be odd.

The set of Supplementary Equations~\eqref{systeff1}-\eqref{systeff2} corresponds to the Su-Schrieffer-Heeger (SSH) model~\cite{Su} with two alternating tunneling amplitudes $j$ and $j+P$, which is known to be the simplest one-dimensional topological model. The only difference from the canonical SSH model is the interaction-induced detuning of the edge sites by $j$ captured by Supplementary Equations~\eqref{boundeff1}, \eqref{boundeff2}.

Solving the system of Supplementary Equations~\eqref{systeff1}-\eqref{systeff2} together with boundary conditions Supplementary Equations~\eqref{boundeff1}-\eqref{boundeff2}, we find two states with the localization ratio $z=\beta_{mm}/\beta_{m-2,m-2}$ given by
\begin{equation}\label{LocLeftEdge}
z_{1,2}=\frac{j+P}{2\,j^3}\,\left[2j\,P+P^2\pm\sqrt{(2jP+P^2)^2+4\,j^4}\right]\:.
\end{equation}
Edge-localized states correspond to $|z|<1$. The energies of these states read:
\begin{equation}\label{EnLeftEdge}
\eps_{1,2}=2U+j-\frac{1}{2j}\,\left[2jP+P^2\pm\sqrt{(2jP+P^2)^2+4j^4}\right]\:.
\end{equation}
Supplementary Equation~\eqref{LocLeftEdge} shows that the higher-energy edge state $\eps_2$ is localized for any $P\not=0$, while the lower-energy state $\eps_1$ disappears when
\begin{equation}\label{LocCondition}
P>0 \mspace{6mu} \text{or} \mspace{6mu} P<-2J^2/U\:.
\end{equation}
In our case with $U/J=7.09$ and $P/J=-4.18$ the latter condition is fulfilled, which means that only the higher-energy state $\eps_2$ persists. Note also that the condition Supplementary Equation~\eqref{LocCondition} is equivalent to $|j+P|>j$ which ensures that the site $(N,N)$ is the weak link edge and the respective edge state can be interpreted as the  topological state inherent to SSH model.

%%%%%%%%%%%%%%%%%%%%%%%%%%%%%%%%%%%%%%%%%%%%%%%%%%%%%%%%%%%%%%%%%%%%%%%%%%%
\section{Supplementary Note 2~-- Eigenmode tomography}\label{sec:Tomography}

In an experimental situation, we can examine the excitation of the system applying voltage to various sites. However, we have no direct access to the system eigenmodes and associated probability distributions, which eventually limits our possibilities to observe topological physics in the designed setup.

To overcome this limitation, we elaborate on the procedure of eigenmode reconstruction (tomography) following the proposals of our recent work~\cite{Gorlach-2018}. The key idea of this method is to collect the information on system excitation when the voltage is applied to various pairs of sites $(m,n)$ symmetrically with respect to the diagonal. The eigenmode profile is then recovered as a sum of squares of voltages in all nodes of the circuit plotted as a function of the feeding point $(m,n)$. In this section, we provide a summary of the proposed technique highlighting further applications of the developed method and its applicability to a wide class of systems described by tight-binding equations.

We consider a two-dimensional system depicted in Supplementary Figure~\ref{fig:Setup}. We describe the behavior of the system under external coherent driving with the following semi-empirical coupled mode equations:
\begin{equation}\label{Driven}
\omega\,c_{mn}=\sum\limits_{m',n'}\,H_{mn,m'n'}\,c_{m'n'}-i\gamma\,c_{mn}-i\kappa\,E_{mn}\:.
\end{equation}
Here $H_{mn,m'n'}$ is the Hamiltonian of a closed system which is assumed Hermitian, $\gamma$ quantifies the dissipation rate, and $E_{mn}$ stands for the  applied electromotive force (or injected current, depending on the  interpretation of $c_{mn}$ coefficients) at frequency $\omega$. In an actual experimental situation, we can apply different drivings $E_{mn}$ at different frequencies and measure the resultant intensity distribution $|c_{mn}|^2$. Our goal is to elaborate a protocol to reconstruct the profile of the system eigenmode $\beta_{mn}^{(k)}$ corresponding to the eigenfrequency $\omega_k$ which is the solution of the eigenvalue problem
\begin{equation}\label{Eigenmode}
\omega_k\,\beta_{mn}^{(k)}=\sum\limits_{m',n'}\,H_{mn,m'n'}\,\beta_{m'n'}^{(k)}\:.
\end{equation}

\begin{figure}
\begin{center}
\includegraphics[width=0.4\linewidth]{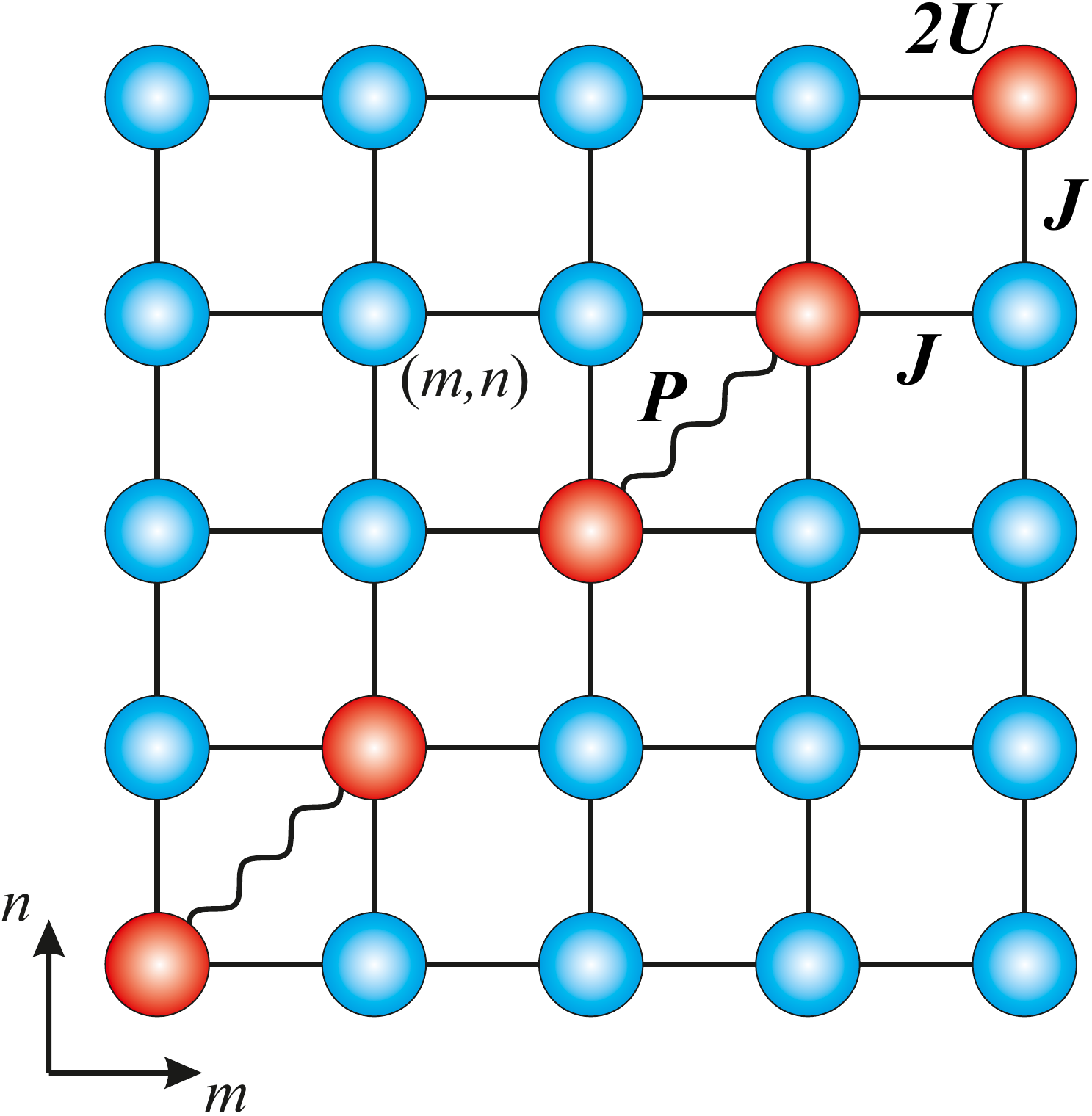}
\caption{Illustration of a two-dimensional setup used for the procedure of eigenmode tomography. For this technique, specific type of tight-binding model is not essential.
\label{fig:Setup}}
\end{center}
\end{figure}

Since the Hamiltonian of the closed system is Hermitian, the eigenfrequencies $\omega_k$ are purely real and the eigenmodes $\beta_{mn}^{(k)}$ are mutually orthogonal, i.e.
\begin{equation}\label{Orthogonality}
\sum\limits_{m,n}\,\beta_{mn}^{(k)}\,\left[\beta_{mn}^{(k')}\right]^*=\delta_{kk'}\:.
\end{equation}
Furthermore, the modes $\beta_{mn}^{(k)}$ form a complete basis and the coefficients $c_{mn}$ can be expanded as
\begin{equation}\label{Expansion}
c_{mn}=\sum\limits_k\,\alpha_k\,\beta_{mn}^{(k)}\:.
\end{equation}
Putting the expansion Supplementary Equation~\eqref{Expansion} into Supplementary Equation~\eqref{Driven}, we obtain:
\begin{gather}
\omega\,\sum\limits_k\,\alpha_k\,\beta_{mn}^{(k)}=\sum\limits_{m',n',k}\,H_{mn,m'n'}\,\alpha_k\,\beta_{m'n'}^{(k)}-i\gamma\,\sum\limits_k\,\alpha_k\,\beta_{mn}^{(k)}-i\kappa\,E_{mn}=\\ \overset{\eqref{Eigenmode}}{=}\sum\limits_{k}\,\alpha_k\,(\omega_k-i\gamma)\,\beta_{mn}^{(k)}-i\kappa\,E_{mn}\:.
\end{gather}
Using the orthogonality property Supplementary Equation~\eqref{Orthogonality}, we recover that
\begin{equation}\label{Superposition}
\left(\omega-\omega_k+i\,\gamma\right)\,\alpha_k=-i\kappa\,\sum\limits_{m,n}\,E_{mn}\,\left[\beta_{mn}^{(k)}\right]^*\:.
\end{equation}
Physically, Supplementary Equation~\eqref{Superposition} expresses an intuitive fact that the larger the overlap of the driving profile with the eigenmode, the larger is the contribution of the eigenmode into the driven-dissipative system stationary state. On the other hand, the closer is the driving frequency to the resonance, the larger is the contribution of a given eigenmode to the driven-dissipative system stationary state. The coefficients $c_{mn}$ are expressed as follows:
\begin{equation}\label{Cmn1}
c_{mn}\overset{\eqref{Expansion}}{=}\sum\limits_k\,\alpha_k\,\beta_{mn}^{(k)}=-i\kappa\,\sum\limits_{m',n',k}\,E_{m'n'}\,\frac{\left[\beta_{m'n'}^{(k)}\right]^*\,\beta_{mn}^{(k)}}{\omega-\omega_k+i\gamma}\:.
\end{equation}
At this point we assume that the driving profile (a) includes only one or two points; (b) is symmetric with respect to the diagonal $m=n$. In the other words,
\begin{equation}
E_{mn}=\frac{E_0}{2}\,\left[\delta_{mp}\,\delta_{nq}+\delta_{mq}\,\delta_{np}\right]\:,
\end{equation}
where $p$ and $q$ provide the coordinates of the excitation point(s). This assumption immediately simplifies the sum. Since the system is symmetric with respect to the diagonal, all non-degenerate eigenmodes are either even or odd; degenerate modes can also be enforced to satisfy this condition by choosing suitable linear combinations. Next, all odd modes have zero overlaps with symmetric pumping profile and they drop out of the sum. The remaining even modes have $\beta_{pq}^{(k)}=\beta_{qp}^{(k)}$. Hence,
\begin{equation}\label{Cmn2}
c_{mn}=-i\kappa\,E_0\,\sum_k\,\frac{\beta_{mn}^{(k)}\,\left[\beta_{pq}^{(k)}\right]^*}{\omega-\omega_k+i\gamma}\:.
\end{equation}
Next, we construct the following quantity:
\begin{gather}\label{Intensity}
\mathfrak{I}(p,q)=\sum\limits_{m,n}\,|c_{mn}|^2=\kappa^2\,|E_0|^2\,\sum\limits_{m,n,k,k'}\,\frac{\beta_{mn}^{(k)}\,\left[\beta_{mn}^{(k')}\right]^*\,\left[\beta_{pq}^{(k)}\right]^*\,\beta_{pq}^{(k')}}{(\omega-\omega_k+i\gamma)\,(\omega-\om_{k'}-i\gamma)}\notag\\
\overset{\eqref{Orthogonality}}{=}\kappa^2\,|E_0|^2\,\sum\limits_{k,k'}\,\frac{\delta_{kk'}\,\left[\beta_{pq}^{(k)}\right]^*\,\beta_{pq}^{(k')}}{(\omega-\omega_k+i\gamma)\,(\omega-\om_k'-i\gamma)}=\kappa^2\,|E_0|^2\,\sum\limits_k\,\frac{|\beta_{pq}^{(k)}|^2}{(\om-\om_k)^2+\gamma^2}\:.
\end{gather}
Thus, for symmetric pumping into $(p,q)$ and $(q,p)$ with the same amplitudes $E_0/2$ and $E_0/2$ and equal phases the sum of squares of field amplitudes in all sites of the system reads:
\begin{equation}\label{Iexpression}
\mathfrak{I}(p,q)\equiv \sum\limits_{m,n}\,|c_{mn}|^2=\kappa^2\,|E_0|^2\,\sum\limits_k\,\frac{|\beta_{pq}^{(k)}|^2}{(\om-\om_k)^2+\gamma^2}\:,
\end{equation}
where the sum extends only over the symmetric eigenmodes. If the driving frequency is close enough to the eigenfrequency $\om_k$, the obtained distribution $\mathfrak{I}(p,q)$ will closely resemble the distribution $|\beta_{pq}^{(k)}|^2$, which is actually the eigenmode intensity (or two-photon probability distribution in the original 1D two-particle problem).

The outlined eigenmode reconstruction protocol works especially well once the spectral distance from the given mode to the rest of the modes exceeds the dissipation rate $\gamma$. In such a case, reconstruction of the eigenmode will be very precise as can be seen from the comparison made in the main text. This technique is also useful for the group of modes, e.g. doublon band, well-separated from the rest of the modes.

Note also that the developed eigenmode reconstruction technique is very general and applicable to a wide range of physical systems. In this work, we apply this protocol to LC circuits. Further identification of driven-dissipative model Supplementary Equation~\eqref{Driven} with Kirchhoff's equations for the electric circuit is provided in Supplementary Notes 3-4.
%~\ref{sec:Mapping1}-\ref{sec:Mapping2}.

%%%%%%%%%%%%%%%%%%%%%%%%%%%%%%%%%%%%%%%%%%%%%%%%%%%%%%%%%%%%%%%%%%%%%%%%%%
\section{Supplementary Note 3~-- Mapping of tight-binding model onto the topolectrical circuit: ideal lossless case}\label{sec:Mapping1}

As indicated in the article main text, the two-particle one-dimensional quantum problem with extended Bose-Hubbard Hamiltonian is described by the following system of tight-binding equations:
\begin{gather}
\left(\eps-2U\right)\,\beta_{2m+1,2m+1}=-2J\,\beta_{2m,2m+1}-2J\,\beta_{2m+1,2m+2}+P\,\beta_{2m+2,2m+2}\:,\label{Bulk1}\\
\left(\eps-2U\right)\,\beta_{2m,2m}=-2J\,\beta_{2m-1,2m}-2J\,\beta_{2m,2m+1}+P\,\beta_{2m-1,2m-1}\:,\label{Bulk2}\\
\eps\,\beta_{mn}=-J\,\beta_{m-1,n}-J\,\beta_{m+1,n}-J\,\beta_{m,n-1}-J\,\beta_{m,n+1}\mspace{10mu} (m\not=n)\:.\label{Bulk3}
\end{gather}
with $\beta_{mn}=\beta_{nm}$ and open boundary conditions at the edges and corners:
\begin{gather}
(\eps-2U)\,\beta_{11}=-2J\,\beta_{12}+P\,\beta_{22}\:,\label{Corner1}\\
\eps\,\beta_{1m}=-J\,\left[\beta_{1,m-1}+\beta_{1,m+1}+\beta_{2,m}\right]\:,\label{Edge}\\
(\eps-2U)\,\beta_{NN}=-2J\,\beta_{N,N-1}\:,\label{Corner2}
\end{gather}
where $N$ is the size of the system.

\begin{figure}
\begin{center}
\includegraphics[width=0.7\linewidth]{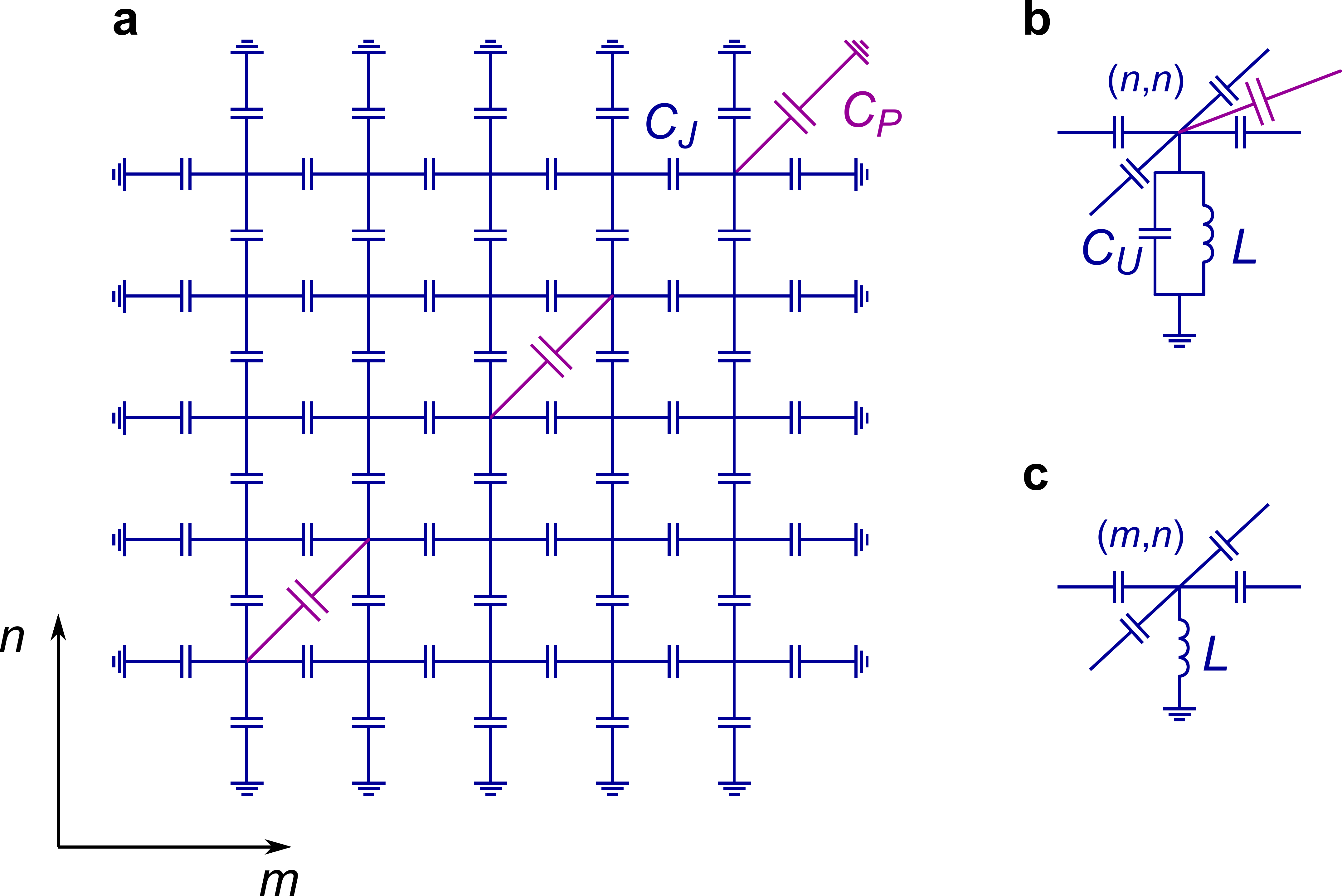}
\caption{a, Implementation of a 2D system emulating extended Bose-Hubbard model, top view. b, Side view of a diagonal node in the 2D system. c, Side view of an off-diagonal node in the 2D system. \label{fig:Scheme}}
\end{center}
\end{figure}

In this Supplementary Note, we analyze the 2D system in Fig.~\ref{fig:Scheme}a with Kirchhoff's circuit laws and express the parameters of the Hamiltonian $U$ and $P$ in terms of the circuit parameters. In our analysis we assume that the voltages in the circuit nodes $U_{mn}\propto e^{-i\om t}$, and hence $\ddot{U}_{mn}=-\om^2\,U_{mn}$. We also assume $J=1$ and denote

\begin{gather}
\om_0^2=\frac{1}{L\,C_j}\:, \mspace{10mu} \eps=\frac{\om_0^2}{\om^2}-4\:,\label{Om0-Eps}\\
U=\frac{C_P+C_U}{2\,C_j}\:, \mspace{10mu} P=-\frac{C_P}{C_j}\:.\label{Udef-Pdef}
\end{gather}

To establish one-to-one correspondence of the initial tight-binding problem with LC circuit, we analyze several representative situations.

	{\it 1. Site $(m,n)$ with $m\not=n$, not at the edge of the system.} From the first Kirchhoff's law we get
\begin{equation}
\frac{U_{mn}}{L}-\om^2\,C_j\,\left[\left(U_{mn}-U_{m,n+1}\right)+\left(U_{mn}-U_{m,n-1}\right)+\left(U_{mn}-U_{m+1,n}\right)+\left(U_{mn}-U_{m-1,n}\right)\right]=0
\end{equation}
or with designations Supplementary Equations~\eqref{Om0-Eps}-\eqref{Udef-Pdef}
\begin{equation}
\eps\,U_{mn}=-U_{m,n+1}-U_{m,n-1}-U_{m+1,n}-U_{m-1,n}\:,
\end{equation}
which is consistent with Supplementary Equation \eqref{Bulk3}.

	{\it 2. Site $(1,n)$ with $n\not=1$, at the edge of the system.}
\begin{equation}\label{SiteCircuit}
\frac{U_{1n}}{L}-\om^2\,C_j\,\left[\left(U_{1n}-U_{1,n-1}\right)+\left(U_{1n}-U_{1,n+1}\right)+\left(U_{1n}-U_{2n}\right)+U_{1n}\right]=0\:,
\end{equation}
where the last term in the square bracket is associated with extra capacitance $C_j$ connected to the ground. Supplementary Equation~\eqref{SiteCircuit} can be rearranged to yield
\begin{equation}
\eps\,U_{1n}=-U_{1,n-1}-U_{1,n+1}-U_{2,n}\:,
\end{equation}
which reproduces open boundary condition for the tight-binding model Supplementary Equation~\eqref{Edge}. The equations for the sites $(m,1)$, $(m,N)$ and $(N,n)$ (not at the corner) where $N$ is the system size are completely analogous.

	{\it 3. Site $(2m+1,2m+1)$ at the diagonal of the system, not the corner one.}
\begin{equation}
\begin{split}
&\frac{U_{2m+1,2m+1}}{L}-\om^2\,C_j\,\left[\left(U_{2m+1,2m+1}-U_{2m,2m+1}\right)+\left(U_{2m+1,2m+1}-U_{2m+2,2m+1}\right)\right.\\
&+\left(U_{2m+1,2m+1}-U_{2m+1,2m}\right)+\left.\left(U_{2m+1,2m+1}-U_{2m+1,2m+2}\right)\right]-\\
&-\om^2\,C_p\,\left(U_{2m+1,2m+1}-U_{2m+2,2m+2}\right)-\om^2\,C_U\,U_{2m+1,2m+1}=0\:,
\end{split}
\end{equation}	
which can be rewritten as
\begin{equation}
\left(\eps-2U\right)\,U_{2m+1,2m+1}=-U_{2m,2m+1}-U_{2m+2,2m+1}-U_{2m+1,2m}-U_{2m+1,2m+2}+PU_{2m+2,2m+2}
\end{equation}
and coincides with Supplementary Equation~\eqref{Bulk1} provided the pattern of voltages is symmetric, i.e. $U_{2m,2m+1}=U_{2m+1,2m}$ and $U_{2m+2,2m+1}=U_{2m+1,2m+2}$.

	{\it 4. Site $(2m,2m)$ at the diagonal of the system}
\begin{equation}
\begin{split}
&\frac{U_{2m,2m}}{L}-\om^2\,C_j\,\left[\left(U_{2m,2m}-U_{2m,2m+1}\right)+\left(U_{2m,2m}-U_{2m,2m-1}\right)\right.+\\
&\left.+\left(U_{2m,2m}-U_{2m+1,2m}\right)+\left(U_{2m,2m}-U_{2m-1,2m}\right)\right]-\\
&-\om^2\,C_p\,\left(U_{2m,2m}-U_{2m-1,2m-1}\right)-\om^2\,C_U\,U_{2m,2m}=0\:,
\end{split}
\end{equation}
which can be recast in the form
\begin{equation}
\left(\eps-2U\right)\,U_{2m,2m}=-U_{2m,2m+1}-U_{2m,2m-1}-U_{2m+1,2m}-U_{2m-1,2m}  +P\,U_{2m-1,2m-1}
\end{equation}
in agreement with Supplementary Equation~\eqref{Bulk2}.

	{\it 5. Site $(1,1)$ at the corner of the system.} For this corner site, we include two capacitances $C_j$ to the ground to compensate the absence of the two neighbors.
\begin{equation}
\frac{U_{11}}{L}-\om^2\,C_j\left[\left(U_{11}-U_{12}\right)+\left(U_{11}-U_{21}\right)+2\,U_{11}\right]-\om^2\,C_p\,\left(U_{11}-U_{22}\right)-\om^2\,C_U\,U_{11}=0\:,
\end{equation}
which is equivalent to
\begin{equation}
(\eps-2U)\,U_{11}=-U_{12}-U_{21}+P\,U_{22}
\end{equation}
exactly reproducing an open boundary condition Supplementary Equation~\eqref{Corner1} for the corner of the system.

	{\it 6. Site $(N,N)$ at the corner of the system ($N$ is assumed to be odd).} In contrast to $(1,1)$ site, besides two capacitances $C_j$ connected to the ground, we also include an extra capacitance $C_P$ connected to the ground:
\begin{equation}
\begin{split}
&\frac{U_{NN}}{L}-\om^2\,C_j\,\left[\left(U_{NN}-U_{N-1,N}\right)+\left(U_{NN}-U_{N,N-1}\right)+2\,U_{NN}\right]-\\
&-\om^2\,C_p\,U_{NN}-\om^2\,C_U\,U_{NN}=0\:,
\end{split}
\end{equation}
which can be rearranged to yield
\begin{equation}
(\eps-2U)\,U_{NN}=-U_{N-1,N}-U_{N,N-1}
\end{equation}
in agreement with open boundary condition for the corner of the system, Supplementary Equation~\eqref{Corner2}.

Hence, the proposed experimental setup allows us to emulate two-body physics in Bose-Hubbard model, but with several constraints on parameters that enter the Bose-Hubbard Hamiltonian: (i) $P$ is always negative, cf. Supplementary Equation~\eqref{Udef-Pdef}; (ii) $U>|P|/2$, i.e. we cannot realize the regime of too weak on-site interactions; (iii) to emulate boson states, the pattern of voltages in a circuit should be symmetric with reasonable accuracy.

%%%%%%%%%%%%%%%%%%%%%%%%%%%%%%%%%%%%%%%%%%%%%%%%%%%%%%%%%%%%%%%%%%%%%%%%%%
\section{Supplementary Note 4~-- Analysis of topolectrical circuit with realistic losses}\label{sec:Mapping2}

As we have demonstrated, the lossless 2D topolectrical circuit corresponds precisely to the Bose-Hubbard model. However, realistic circuits necessarily have losses, and in this section, we examine excitation of the system taking the effect of loss into account. We assume that the system is excited with a current source and current is pumped into one or several lattice sites.  We aim to compare the governing equations with the simple driven-dissipative model outlined in Supplementary Note 2, which is described by the equation
\begin{equation}\label{DrivenDiss}
\om\,c_{mn}=\sum\limits_{m',n'}H_{mn,m'n'}\,c_{m'n'}-i\,\gamma\,c_{mn}-i\kap\,E_{mn}\:,
\end{equation}
where $\omega$ is a driving frequency, $c_{mn}$ describe the stationary state of the system, $\gamma$ is the dissipation rate and $\kap$ is the coupling coefficient. 

In circuit analysis, we assume $e^{-i\om t}$ time dependence of voltages, which yields the following impedances:
\begin{gather}
Z_L=-i\om L+R_L\:,\mspace{10mu} Z_C=\frac{1}{-i\om\, C_j}+R_C\:,\\
Z_P=\frac{1}{-i\om\, C_P}+R_P\:, \mspace{10mu} Z_U=\frac{1}{-i\om\,C_U}+R_U\:.
\end{gather}
For sufficiently small losses, the ratios of impedances read
\begin{equation}
\frac{Z_C}{Z_L} =-\frac{\om_0^2}{\om^2} +i \left(\frac{R_L\,\om_0^2}{L\,\om^3}+\frac{R_C}{L\,\om}\right)\:,
\end{equation}
\begin{gather}
\frac{Z_C}{Z_P} =\frac{C_P}{C_j}+i\,\om\,\frac{C_P}{C}\,\left(R_P\,C_P-R_C\,C_j\right)\:,\\
\frac{Z_C}{Z_U} =\frac{C_U}{C_j}+i\,\om\,\frac{C_U}{C_j}\,\left(R_U\,C_U-R_C\,C_j\right)\:.
\end{gather}
Similarly to Supplementary Note 3, we analyze several characteristic situations:

{\it 1. Site $(m,n)$ with $m\not=n$, not at the edge of the system.} First Kirchhoff's law now yields:
\begin{equation}
\frac{U_{mn}}{Z_L} +\frac{1}{Z_C}\,\left(4U_{mn}-U_{m,n-1}-U_{m,n+1}-U_{m-1,n}-U_{m+1,n}\right) =I_{mn}
\end{equation}
which can be rearranged as
\begin{equation}
\eps\,U_{mn}=-U_{m,n-1}-U_{m,n+1}-U_{m-1,n}-U_{m+1,n}+i\,\gamma\,U_{mn}-Z_C\,I_{mn}\:,
\end{equation}
where $\gamma$ is a frequency-dependent damping
\begin{equation}\label{GammaDef}
\gamma=\frac{R_L\,\om_0^2}{L \om^3}+\frac{R_C}{L\,\om}\:.
\end{equation}
Note also the sign in front of $\gamma$ in Supplementary Equation~\eqref{GammaDef}: it is different from the sign in Supplementary Equation~\eqref{DrivenDiss}. This happens due to the definition of ``energy'' variable Supplementary Equation~\eqref{Om0-Eps} which ensures that imaginary parts of $\om$ and $\eps$ have opposite signs:
\begin{equation}
\om''=-\frac{\om_0}{2}\,\frac{\eps''}{\left(\eps'+4\right)^{3/2}}\:.
\end{equation}
Since in the dissipative case eigenmode frequency has negative imaginary part, ``energy variable'' $\eps$ should have positive imaginary part.

{\it 2. Site $(1,n)$ with $n\not =1$, at the edge of the system.} In a similar way we derive an equation
\begin{equation}
\frac{U_{1n}}{Z_L}+\frac{U_{1n}}{Z_C}+\frac{1}{Z_C}\,\left(3\,U_{1n}-U_{1,n-1}-U_{1,n+1}-U_{2n}\right) =I_{1n}\:,
\end{equation}
which yields
\begin{equation}
\eps\,U_{1n}=-U_{1,n-1}-U_{1,n+1}-U_{2n}+i\,\gamma\, U_{1n}-Z_C\,I_{1n}\:.
\end{equation}

{\it 3. Site $(2m+1,2m+1)$ at the diagonal of the system, not the corner one.}
\begin{equation}
\begin{split}
&\frac{U_{2m+1,2m+1}}{Z_L}+\frac{U_{2m+1,2m+1}}{Z_U}+\frac{U_{2m+1,2m+1}-U_{2m+2,2m+2}}{Z_P} +\\
&+\frac{1}{Z_C}\,\left(4\,U_{2m+1,2m+1}-U_{2m,2m+1}-U_{2m+2,2m+1}-U_{2m+1,2m}-U_{2m+1,2m+2}\right) =I_{2m+1,2m+1}
\end{split}
\end{equation}
This yields an equation
\begin{equation}
\begin{split}
&(\eps-2U) \,U_{2m+1,2m+1}=-U_{2m,2m+1}-U_{2m+2,2m+1}-U_{2m+1,2m}-U_{2m+1,2m+2}+\\
&+i\gamma'\,U_{2m+1,2m+1}+P'\,U_{2m+2,2m+2}-Z_C\, I_{2m+1,2m+1}\:,
\end{split}
\end{equation}
where for the diagonal sites
\begin{gather}
\gamma'=\gamma+\om\,\frac{C_P}{C_j}\,\left(R_P\,C_P-R_C\,C_j\right)+\om\,\frac{C_U}{C_j}\,\left(R_U \,C_U-R_C\,C_j\right)\:,\\
P'=P-i\om\,\frac{C_P}{C_j}\,\left(R_P\,C_P-R_C\,C_j\right)\:.
\end{gather}
Thus, the diagonal elements have extra loss in tunneling $P'$ and on-site losses $\gamma'$ stemming from the  insertion of additional elements. This problem, however, can be circumvented by requiring that
\begin{equation}\label{AddCons}
R_C\,C_j=R_P\,C_P=R_U\,C_U
\end{equation}
in which case all elements (in the bulk and at the diagonal) are characterized by the same magnitude of loss, though depending on frequency. However, even if Supplementary Equation~\eqref{AddCons} is not fulfilled, doublon bands will remain almost unaffected, since the maximal voltages are expected at the diagonal sites only.

{\it 4. Site $(2m,2m)$ at the diagonal of the system.}
\begin{equation}
\begin{split}
&\frac{U_{2m,2m}}{Z_L} +\frac{U_{2m,2m}}{Z_U} +\frac{U_{2m,2m}-U_{2m-1,2m-1}}{Z_P} +\\
&+\frac{1}{Z_C}\,\left(4U_{2m,2m}-U_{2m-1,2m}-U_{2m+1,2m}-U_{2m,2m-1}-U_{2m,2m+1}\right)=I_{2m,2m}
\end{split}
\end{equation}
which yields
\begin{equation}
\begin{split}
&(\eps-2U)\, U_{2m,2m}=-U_{2m-1,2m}-U_{2m+1,2m}-U_{2m,2m-1}-U_{2m,2m+1}+\\
&+i\gamma'\,U_{2m,2m}+P'\,U_{2m-1,2m-1}-Z_C\,I_{2m,2m}\:.
\end{split}
\end{equation}

{\it 5. Site $(1,1)$ at the corner of the system.}
\begin{equation}
\left(Z_L^{-1}+Z_U^{-1}+2\,Z_C^{-1}\right)\,U_{11}+\frac{1}{Z_C}\,\left(2U_{11}-U_{12}-U_{21}\right)+\frac{1}{Z_P}\,\left(U_{11}-U_{22}\right)=I_{11}\:,
\end{equation}
which yields
\begin{equation}
(\eps-2U)\, U_{11}=i\gamma'\,U_{11}-U_{12}-U_{21}+P'\,U_{22}-Z_C\,I_{11}\:.
\end{equation}

{\it 6. Site $(N,N)$ at the corner of the system [$N$ is assumed to be odd].}
\begin{equation}
\left(Z_L^{-1}+Z_U^{-1}+Z_P^{-1}+2\,Z_C^{-1}\right)\,U_{NN}+\frac{1}{Z_C}\,\left(2\,U_{NN}-U_{N-1,N}-U_{N,N-1}\right)=I_{NN}
\end{equation}
which can be transformed as
\begin{equation}
(\eps-2U)\, U_{NN}=i\,\gamma'\,U_{NN}-U_{N-1,N}-U_{N,N-1}-Z_C\,I_{NN}\:.
\end{equation}

To summarize, the designed LC circuit is described by the same driven-dissipative system Supplementary Equations~\eqref{DrivenDiss} apart from the following differences:
\begin{itemize}
\item	dissipation coefficient $\gamma$ depends on the frequency of driving $\omega$;
\item	the magnitude of dissipation for diagonal and off-diagonal sites is different unless an additional constraint Supplementary Equation~\eqref{AddCons} is fulfilled;
\item instead of frequency in Supplementary Equation~\eqref{DrivenDiss}, we deal with the auxiliary ``energy variable'' $\eps$ which is inversely proportional to frequency and therefore has a positive imaginary part in the dissipative case.
\item instead of external field $E_{mn}$ we have a combination $Z_C I_{mn}$ which is frequency-dependent.
\end{itemize}
In all other aspects, the standard driven-dissipative model captures all essential features of the proposed topolectrical circuit and hence the tomography technique developed in Supplementary Note 2~\ref{sec:Tomography} can be applied to the designed LC circuit.

%%%%%%%%%%%%%%%%%%%%%%%%%%%%%%%%%%%%%%%%%%%%%%%%%%%%%%%%%%%%%%%%%%%%%%%%%%
\section{Supplementary Note 5~-- Distribution of elements in the experimental setup}\label{sec:Experimental_setup}

In contrast to the idealized model of electric circuit discussed above, the actual values of inductances and capacitances of setup elements slightly fluctuate from one element to another. Hence, the experimental circuit possesses an inherent disorder in the tunneling constants $J$ and $P$ (ascribed to capacitors $C_J$ and $C_P$) as well as in the interaction strength $U$ and on-site resonant frequencies $\omega_0$, depending on capacitors $C_U$ and inductors $L$, respectively. The elements used in our experimental sample are Murata GRM32RR71H105KA01 for $C_J$, Murata GRM31CR61A476ME15L for $C_P$, Murata GRM31CR71C106KAC7 for $C_U$, and Bourns RLB1314-220KL for $L$. In the process of fabrication, we created a detailed map of elements that allows us to  determine the precise value of the given circuit bond. The distributions of the element values are shown in  Supplementary Figure~\ref{fig:Elements_Distribution}. As seen from the histograms, actual mean values in the fabricated circuit are $L = 22.77\mspace{2mu}{\rm \mu H}$, $C_J = 1.0024\mspace{2mu}{\rm \mu F}$, $C_U = 10.031\mspace{2mu}{\rm  \mu F}$, and $C_P = 4.1863\mspace{2mu}{\rm \mu F}$ which are slightly different from the ones provided in specifications. Typical fluctuations of parameters are of the order of 1-2\%. We use these measured mean values for numerical simulations described in the Article main text.

\begin{figure}
\begin{center}
\includegraphics[width=0.8\linewidth]{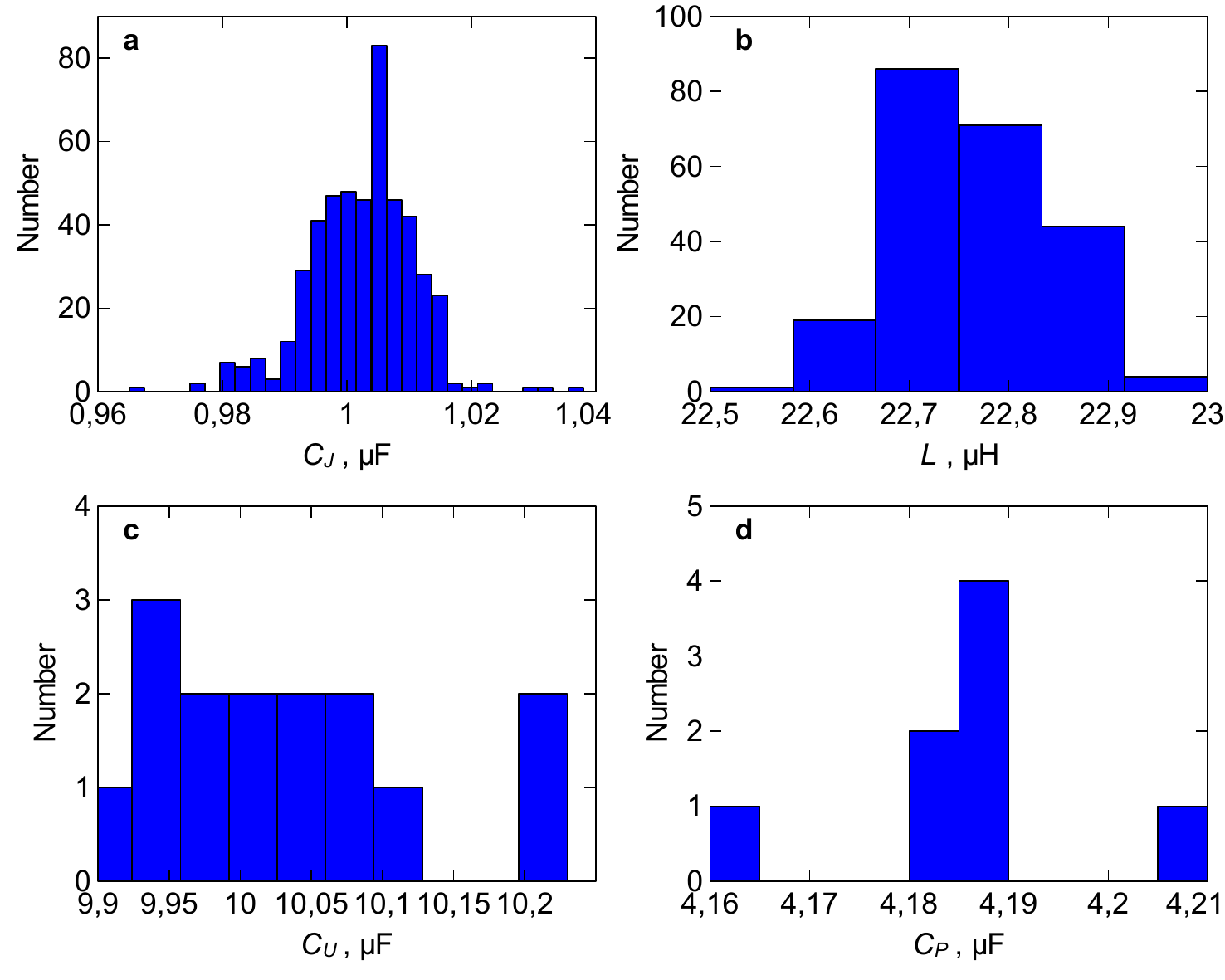}
\caption{Distribution of the lumped element values in the experimental setup for (a) capacitors $C_J$, (b) inductors $L$, (c) $C_U$  and (d) $C_P$. Different number of bins in panels (a) and (b) is due to the different precision of measurements in cases (a-d). Total amounts of corresponding elements in the circuit are 225 for $L$, 480 for $C_J$, 15 for $C_U$, and 8 for $C_P$.\label{fig:Elements_Distribution}}
\end{center}
\end{figure}

%%%%%%%%%%%%%%%%%%%%%%%%%%%%%%%%%%%%%%%%%%%%%%%%%%%%%%%%%%%%%%%%%%%%%%%%%%
\section{Supplementary Note 6~-- Mutual inductance}\label{sec:Mutual_inductance}

Some elements in the experimental setup can have extra parasitic couplings besides the couplings introduced intentionally. This effect is especially pronounced for inductive coils, which have relatively large size compared to the inter-coil distance, as can be seen from Fig.~1 of the article main text. We estimate now the effects of such inductive coupling on the performance of the studied setup.

Incorporating mutual inductance between the coils (grounding elements) into the Kirchhoff's rule for the site $(2m+1,2m+1)$, which we consider as an example, we get:
\begin{equation}
\begin{split}
&(U_{2m+1,2m+1} + U_{\rm ind})(Z_{L}^{-1} + Z_{U}^{-1}) + (U_{2m+1,2m+1} - U_{2m+2,2m+2})Z_{P}^{-1} +\\
&+(4U_{2m+1,2m+1} - U_{2m,2m+1} - U_{2m+2,2m+1} - U_{2m+1,2m} - U_{2m+1,2m+2})Z_{C}^{-1} =\\
&= I_{2m+1,2m+1},
\label{eq:Coupling}
\end{split}
\end{equation}
where $U_{\rm ind}$ denotes the voltage induced in the coil connecting the site $(2m+1,2m+1)$ to the ground by all  surrounding coils. A diagonal capacitor $C_{P}$ is present between the sites $(2m+1,2m+1)$ and $(2m+2,2m+2)$. Taking into account the interaction of the given coil with its nearest neighbors and also with the coils having both coordinates $m$, $n$ shifted by $1$ (diagonal neighbors), we can express this induced voltage as
\begin{equation}
\begin{split}
U_{\rm ind} &= -Z_{M}(I_{2m,2m+1}^{(g)} + I_{2m+2,2m+1}^{(g)} + I_{2m+1,2m}^{(g)} + I_{2m+1,2m+2}^{(g)}) -\\
&- Z_{M}^{\prime}(I_{2m,2m}^{(g)} + I_{2m,2m+2}^{(g)} + I_{2m+2,2m}^{(g)} + I_{2m+2,2m+2}^{(g)}),
\end{split}
\end{equation}
where $I_{m,n}^{(g)}$ is a current through the inductance which connects the site $(m,n)$ to the ground, $Z_{M}$ and $Z_M^{\prime}$ are the impedances corresponding to the mutual inductance between the nearest neighbors and diagonal neighbors, respectively.

In the limit of strong interaction $U\gg 1$, which is the case in the considered model, all voltages are mostly concentrated at the diagonal sites of the circuit in the frequency range of interest. Then, the above expression takes the form
\begin{equation}
\begin{split}
U_{\rm ind} &\approx  - Z_{M}^{\prime}\Big(\frac{U_{2m+1,2m+1} - U_{2m+2,2m+2}}{Z_{P}} + I_{2m+2,2m+2}\Big)- Z_{M}^{\prime}I_{2m,2m},
\end{split}
\end{equation}
where $I_{2m,2m}$ and $I_{2m+2,2m+2}$ denote external currents applied to the corresponding sites. Then, Supplementary Equation~(\ref{eq:Coupling}) reads
\begin{equation}
\begin{split}
&U_{2m+1,2m+1}(Z_{L}^{-1} + Z_{U}^{-1})\Big(1 - \frac{Z_{M}^{\prime}}{Z_{P}}\Big) + (U_{2m+1,2m+1} - U_{2m+2,2m+2})Z_{P}^{-1} =\\
&= I_{2m+1,2m+1} + Z_{M}^{\prime}(Z_{L}^{-1} + Z_{U}^{-1})(I_{2m,2m} + I_{2m+2,2m+2}).
\end{split}
\end{equation}
If only one site of the circuit is driven, which is the case in our measurements, then the second term at the right-hand side vanishes, and we finally obtain the equation
\begin{equation}
    U_{2m+1,2m+1}(Z_{L}^{-1} + Z_{U}^{-1})\Big(1 - \frac{Z_{M}^{\prime}}{Z_{P}}\Big) + (U_{2m+1,2m+1} - U_{2m+2,2m+2})Z_{P}^{-1} = I_{2m+1,2m+1},
\end{equation}
which has the same form as the equation describing the circuit without inductive couplings up to renormalization of the effective model parameters caused by the extra factor $(1 - Z_{M}^{\prime}/Z_{P})$.

%%%%%%%% %%%%%%%%%%%%%%%%%%%%%%%%%%%%%%%%%%%%%%%%%%%%%%%%%%%%%%%%%%%%%%%%%%
\section{Supplementary Note 7~-- Circuit impedance and doublon spectroscopy}\label{sec:Spectroscopy}

\begin{figure}
\begin{center}
\includegraphics[width=0.8\linewidth]{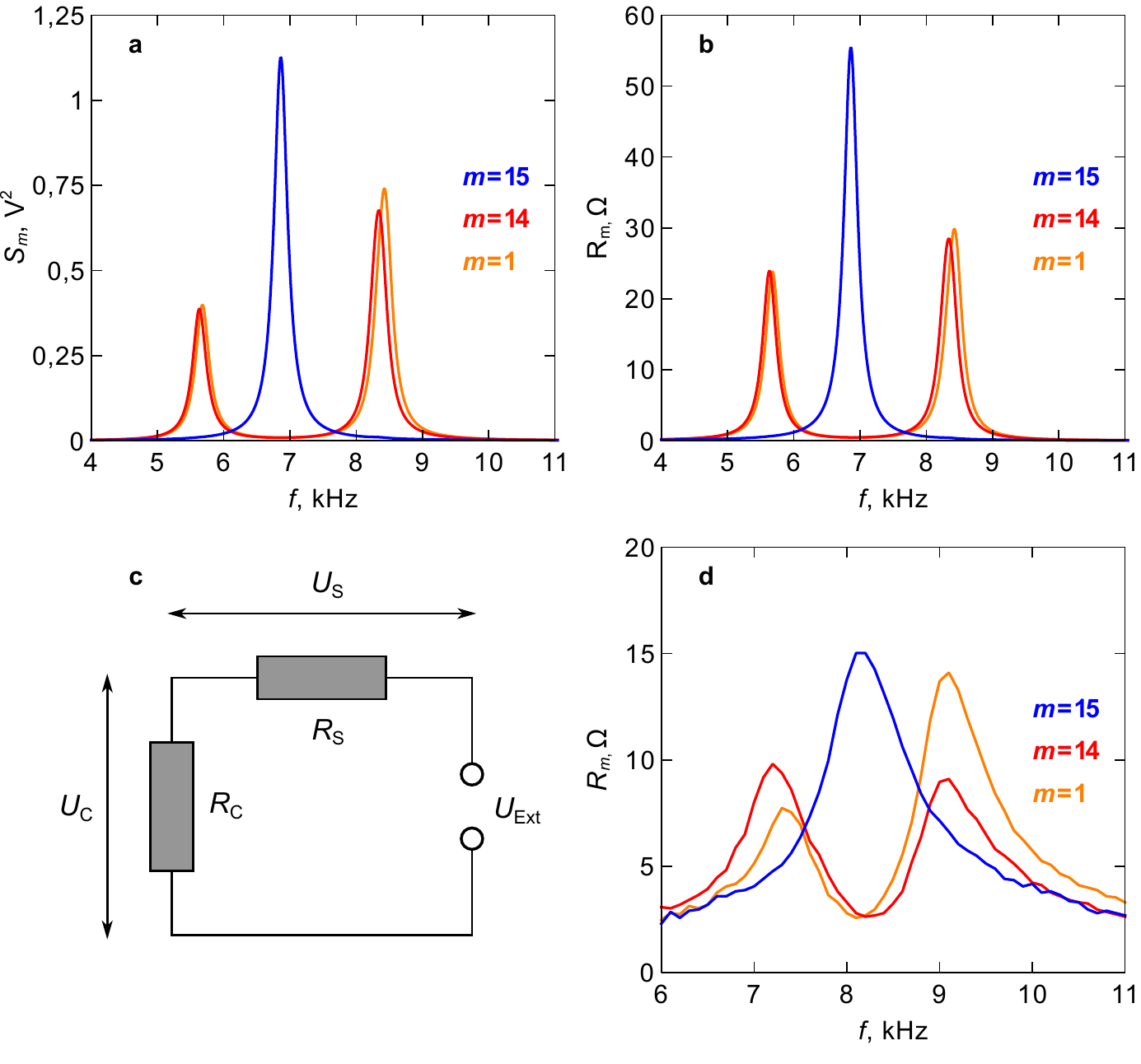}
\caption{(a) Numerically simulated doublon spectroscopy relying on the protocol with quantity $S_m$, Supplementary Equation~(\ref{eq:S_m_Supp}), for $m=1$,$14$ and $15$. (b) Circuit impedance spectroscopy showing the real part $R_m$ of the total impedance between the given site $(m,m)$ and ground. Both panels are calculated for the values of circuit elements taken from the exact map of the experimental setup. (c) Equivalent scheme for the experimental setup including an external voltage source $U_{\rm ext}$ with the resistance $R_s$ and the associated voltage drop $U_s$. $R_c$ denotes the real part of the total impedance between the driven site of the circuit and the ground with the associated voltage drop $U_c$. (d) Experimental results of the circuit impedance spectroscopy.\label{fig:Spectroscopy}}
\end{center}
\end{figure}

To demonstrate doublon states experimentally, we need a technique to distinguish doublon modes from the rest of the modes supported by the two-dimensional sample. In this regard, it is useful to consider the quantity~\cite{Gorlach-2018}
\begin{equation}
    S_m(f) = \sum_{n}|\varphi_{nn}^{mm}|^{2},
    \label{eq:S_m_Supp}
\end{equation}
where $\varphi_{ij}^{mn}$ is the potential at the site $(i,j)$ of the circuit when the external driving voltage at frequency $f$ is applied to the site $(m,m)$. Since doublon modes are characterized by the voltage maxima at the diagonal sites, this quantity exhibits resonant peaks at frequencies of doublon modes. If the external voltage is applied to the site $(N,N)$, where $N=15$ in our case, a single peak corresponding to the doublon edge state is observed. On the other hand, if any other diagonal site is driven, then two peaks centered at frequencies of bulk doublon bands emerge, Supplementary Figure~\ref{fig:Spectroscopy}a.

We now demonstrate that the frequency dependence of the quantity $S_m$ is consistent with the results of the  circuit impedance spectroscopy. Indeed, one can introduce Green's matrix of the circuit $\hat{G}$~\cite{Chung}, which is by definition related to the admittance matrix $\hat{Y}$ introduced in the main text as
\begin{equation}
    G_{mn,m'n'}\, = (Y^{-1})_{mn,m'n'}
\end{equation}
Then, potentials at the nodes of the externally driven circuit are related to the driving current $I_{mn}$ as
\begin{equation}
    \varphi_{mn}\, = \sum\limits_{m',n'}\,G_{mn,m'n'}\,I_{m'n'},
\end{equation}
and therefore $\varphi_{nn}^{mm}=G_{nn,mm}\,I_{\rm ext}$ where $I_{\rm ext}$ is a value of the driving current. Hence, Supplementary Equation~(\ref{eq:S_m_Supp}) takes the form
\begin{equation}
    S_m = I_{\rm ext}^2\, \sum_n |G_{nn,mm}|^2\:.
\end{equation}

At the same time, the characteristic impedance $R_m$ between the given node $(m,m)$ of the circuit and the ground is simply given by the diagonal element of the Green's matrix:
\begin{equation}
   R_{m} = {\rm Re}\{ G_{mm,mm}\}.
\end{equation}
As seen from Supplementary Figure~\ref{fig:Spectroscopy}b, the characteristic impedance of the circuit $R_m$ demonstrates a very similar structure of resonant peaks compared to the quantity $S_m$ for various positions of the driven site $(m,m)$, which highlights the dominant role of diagonal entries of the Green's matrix in our system.

The equivalent scheme of the experimental setup with an external source applied to the diagonal site has the form shown in Supplementary Figure~\ref{fig:Spectroscopy}c. It represents a series connection of the total circuit impedance with the real part $R_c$ and the voltage source equivalent impedance $R_s = 50\mspace{3mu}{\rm \Omega}$. In the experiment, the voltage drop between the given site of the circuit and the ground $U_c$ is measured as a function of the driving frequency $f$. The external voltage $U_{\rm ext}$ is fixed and set to the value 1~V. Then, the circuit impedance $R_c$ can be obtained as
\begin{equation}
    R_c(f) = \frac{U_c(f)}{U_{\rm ext} - U_c(f)}R_s,
\end{equation}
while the external current flowing into the circuit is given by the relation $I_{\rm ext}(f) = U_{\rm ext}/(R_c(f)+R_s)$. All calculations in the article main text are carried on for the constant external current $I_{\rm ext}$ flowing into the system, whereas the experimental data is obtained for the fixed external voltage $U_{\rm ext}$. To make a fair comparison of these results, we multiply the experimental values of $S_m(f)$ by the factor $I^2(f_0)/I^2(f)$ with $f_0=6$~kHz defining a calibrating value of the external current. Experimental results on doublon spectroscopy recalculated in this way are presented in Fig.~2c of the article main text.

%%%%%%%%%%%%%%%%%%%%%%%%%%%%%%%%%%%%%%%%%%%%%%%%%%%%%%%%%%%%%%%%%%%%%%%%%%
\section{Supplementary Note 8~--  Scattering states}\label{sec:Scattering_states}

\begin{figure}
\begin{center}
\includegraphics[width=0.95\linewidth]{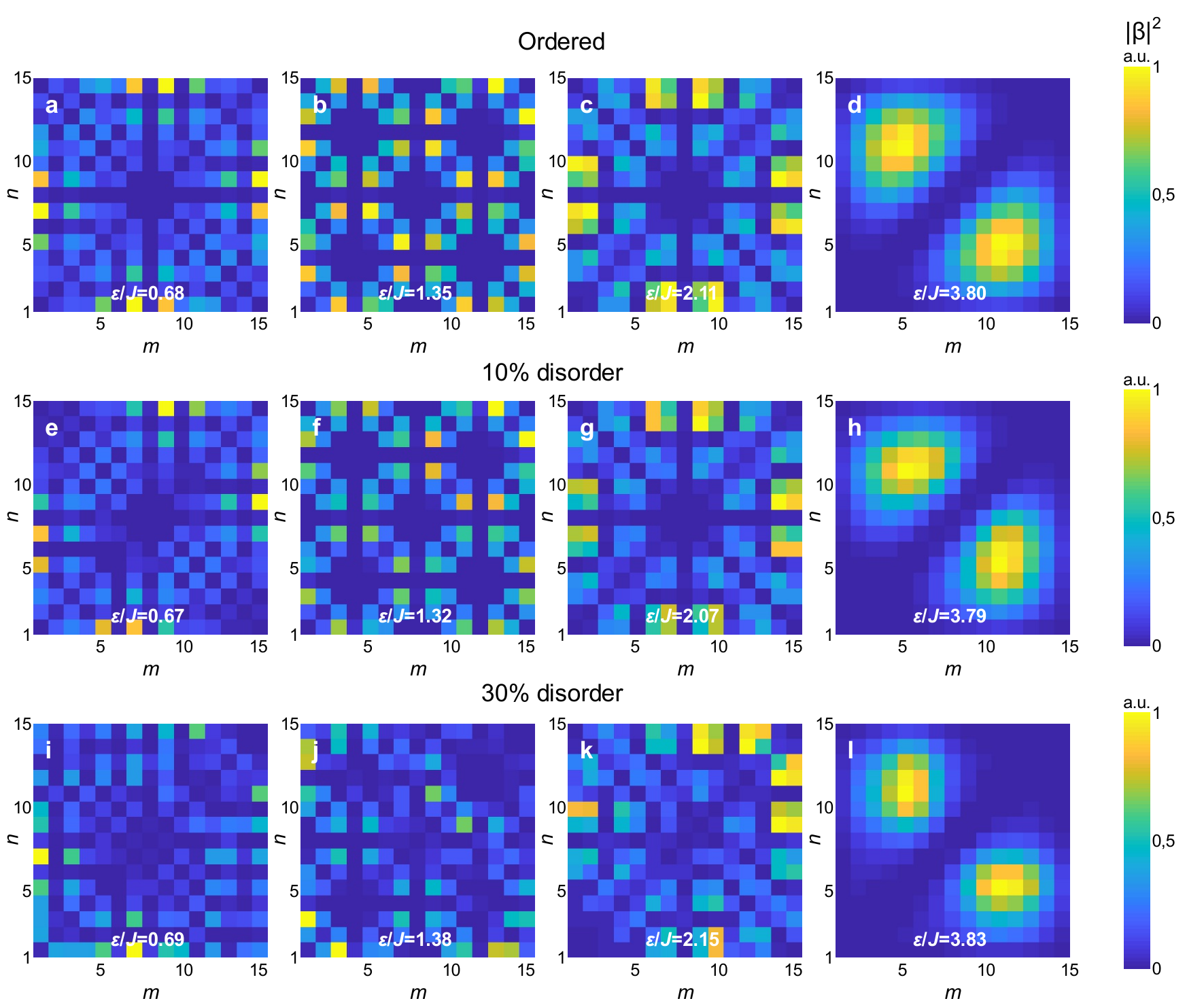}
\caption{Scattering states of photon pair with energies (a) $\varepsilon/J=0.68$, (b) $\varepsilon/J=1.35$, (c) $\varepsilon/J = 2.11$, and (d) $\varepsilon/J = 3.80$ . Color encodes the magnitude of the two-photon probability distribution $|\beta_{mn}|^2$. Panels (a)-(d) are obtained by diagonalization of the tight binding Hamiltonian for the model without disorder. (e-h) Scattering states corresponding to the same eigenmodes as in the top row, but for a uniform $\pm 10$\% disorder in the strength of bonds. (i-l) The same modes as in (e-h), but for a $\pm 30$\% bond disorder.\label{fig:Scattering_states}}
\end{center}
\end{figure}

Besides two bulk doublon bands and doublon edge state, there exists a vast set of the system eigenmodes termed as scattering states. These modes shown by the shaded region in the dispersion (Fig.~2a of the article main text) correspond to such a state of two photons when they are typically located at distinct resonators and possess energy equal to the sum of single-photon energies. Supplementary Figure~\ref{fig:Scattering_states} shows the probability distributions $|\beta_{mn}|^2$ for such states obtained by the diagonalization of the tight-binding Hamiltonian in the absence of dissipation. Here, panels (a-d) correspond to the eigenmodes of an ideal system without any disorder. Introducing the disorder in coupling constants $J$ and $P$ with the uniform distribution within the range $\pm 10$\%, we observe only slight changes in the eigenmode intensity patterns accompanied by weak energy shifts, Supplementary Figure~\ref{fig:Scattering_states}(e-h). However, when the strength of disorder is increased further up to 30\%, quite strong distortions of eigenmode profiles are observed and the energy shifts become comparable with the spectral distance between the spectrally close modes, Supplementary Figure~\ref{fig:Scattering_states}(i-l).

%%%%%%%%%%%%%%%%%%%%%%%%%%%%%%%%%%%%%%%%%%%%%%%%%%%%%%%%%%%%%%%%%%%%%%%%%%
\section{Supplementary Note 9~-- Effects of disorder}\label{sec:Disorder}

To examine the impact of disorder and losses on the results of tomography procedure (Supplementary Note~2), we simulate eigenmode tomography for the designed circuit taking into account Ohmic losses as well as fluctuations in the values of all lumped elements $L$, $C_J$, $C_U$, and $C_P$. For the sake of simplicity, we consider uniformly distributed fluctuations, and Ohmic losses are assumed to be constant and equal to their maximal possible values within the considered spectral range according to elements specifications.

The evolution of low-energy doublon mode, doublon edge state, and high-energy doublon mode with the increase of disorder is shown in panels (a-c), (d-f), and (g-i) of Supplementary Figure~\ref{fig:Disorder}, respectively. It is seen that enhanced localization of doublon modes can be observed even at 10\% disorder, which finally results in the formation of localized states in the bulk of the circuit at 30\% disorder. At the same time, the edge state easily survives 10\% disorder, since it is spectrally well-separated from the bulk doublon bands having a relatively small spatial overlap with them. However, strong 30\% disorder can mix it with bulk doublon states, as seen in Supplementary Figure~\ref{fig:Disorder}f. It should be stressed that the considered levels of disorder in the values of circuit elements exceed those expected for the experimental circuit and the corresponding results are calculated for the illustrative purpose only.

\begin{figure}
\begin{center}
\includegraphics[width=0.8\linewidth]{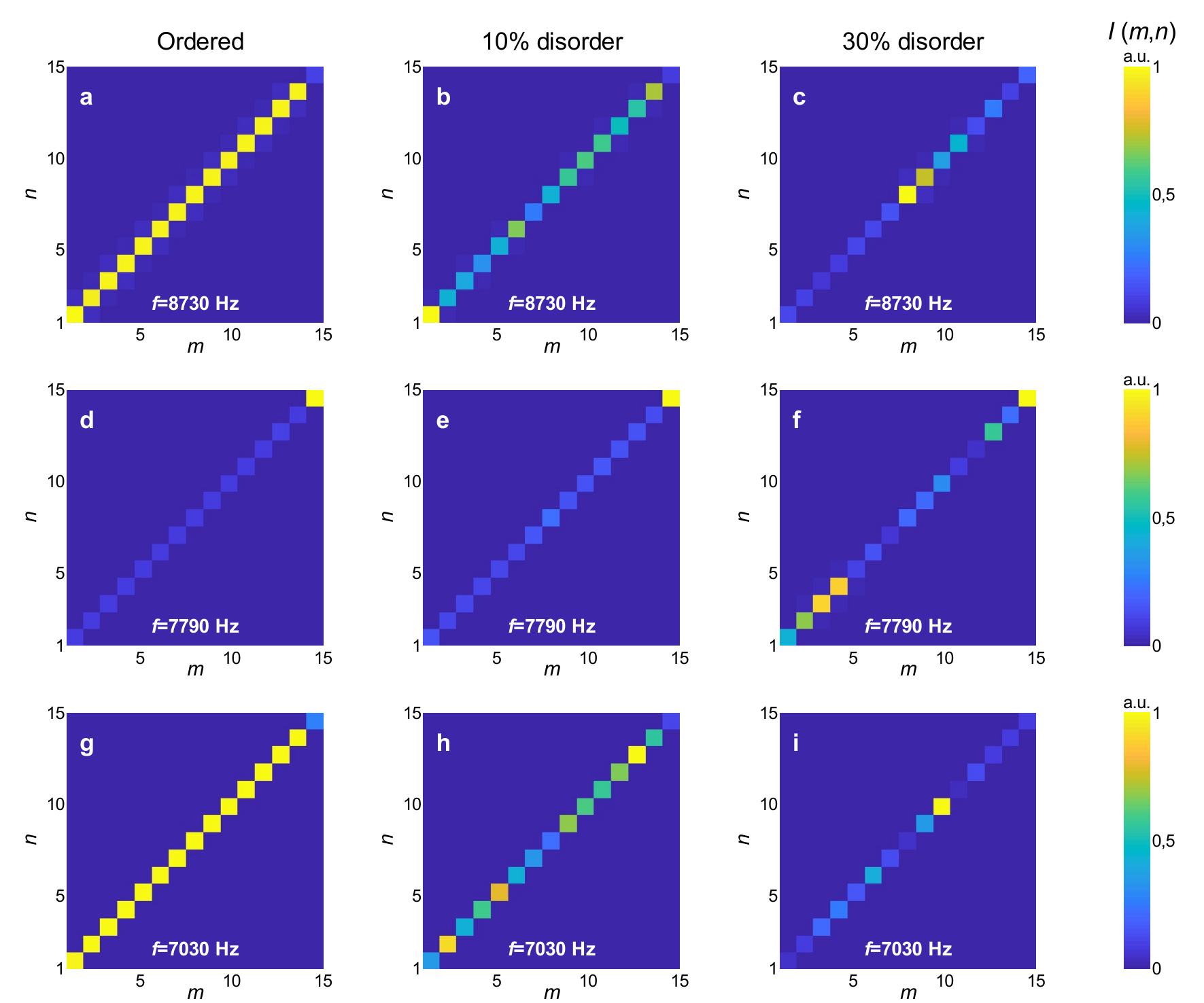}
\caption{Simulation of eigenmode tomography in circuits with various levels of disorder in element values. (a-c) Low-energy (high-frequency) doublon mode, $f=8730$~Hz. (d-f) doublon edge state, $f=7790$~Hz. (g-i): high-energy (low-frequency) doublon mode, $f=7030$~Hz. Simulation of tomography is performed at frequencies corresponding to the peak positions in doublon spectroscopy simulation.\label{fig:Disorder}}
\end{center}
\end{figure}
		
%%%%%%%%%%%%%%%%%%%%%%%%%%%%%%%%%%%%%%%%%%%%%%%%%%%%%%%%%%%%%%%%%%%%%%%%%%%
\section{Supplementary Note 10~-- Evaluation of topological invariant from experimental data}\label{sec:Invariant}

The definition of topological invariant for interacting many-body systems is currently an open problem which is being actively investigated. In our specific case, however, the clue to topological characterization is provided by the fact that the effective doublon Hamiltonian corresponds to that of the Su-Schrieffer-Heeger model once strong interaction regime $U\gg J$ is realized. This is the case for our experimental sample with $U/J=7.09$ and $P/J=-4.18$.

The effective Hamiltonian for doublons written in the basis of isolated resonators' eigenstates takes the form (see Supplementary Note 1 for details):
\begin{equation}\label{BlochHam}
\hat{H}(k)=
\begin{pmatrix}
0 & j_1+j_2\,e^{-ik}\\
j_1+j_2\,e^{ik} & 0
\end{pmatrix}
\:,
\end{equation}
where $j_1=J^2/U$ and $j_2=j_1+P$ are effective tunneling amplitudes for doublons. Hence, for the given Bloch eigenmode the ratio of two components of the doublon wave function $\ket{\psi}=(\psi_{\rm A},\psi_{\rm B})^T$ is given by:
\begin{equation}\label{SublatticeRatio}
\frac{\psi_{\rm A}}{\psi_{\rm  B}}=\frac{j_1+j_2\,e^{-ik}}{\eps(k)}\:.
\end{equation}

We notice that the effective doublon Hamiltonian Supplementary Equation~\eqref{BlochHam} is represented in chiral basis and therefore the winding number is determined by plotting its off-diagonal block $q(k)=j_1+j_2\,e^{-ik}$ on the complex plane~\cite{Ryu}. Furthermore, according to Supplementary Equation~\eqref{SublatticeRatio} $q(k)=\eps(k)\,\psi_{\rm A}/\psi_{\rm B}$, where doublon  energy $\eps(k)$ is purely real. Hence, the winding number can be found by plotting the ratio $\psi_{\rm A}/\psi_{\rm B}$ for a particular Bloch eigenmode.

In an experimental situation, we measure the distribution of voltages at the diagonal of the sample: $U_{\rm A}(n)$ and $U_{\rm B}(n)$. To get a result, corresponding to the given value of wave number $k$, we perform a discrete Fourier transform of those voltages extracting
\begin{gather}
U_{\rm A}(k)=\sum\limits_n\,U_{\rm A}(n)e^{-ikn}\:,\\
U_{\rm B}(k)=\sum\limits_n\,U_{\rm B}(n)e^{-ik(n-1)}
\end{gather}
and then plotting the ratio $U_{\rm A}(k)/U_{\rm B}(k)$ on the complex plane.

Calculating the winding number from experimental data, we assume that the pattern of voltages excited in a topolectrical circuit by the source inserted in the middle of the diagonal resembles the pattern of voltages expected for the eigenmode, which is justified in the case of doublon bands well-separated from the scattering continuum.

\bibliographystyle{naturemag}
\bibliography{TopologicalLib}